\documentclass[a4paper]{jpconf}
\usepackage{graphicx}
\usepackage{amssymb}
\usepackage{epsfig}
\pdfoutput=1

\def\href#1{}
\def\splitJINST#120#2#3#4#5#6#7#8#9{\href{http://www.iop.org/EJ/abstract/1748-0221/#1/#5#6/#4#5#6#7#8#9}
		{20#2#3 {\it JINST }{\bf #1} #4#5#6#7#8#9}}

\newcommand{\mgcs}{\ensuremath{\,\mathrm{mg/cm^{2}}}}

\def\nue{\nu_e}
\def\nutau{\nu_\tau}

\def\numu{\nu_\mu}

\begin{document}
\title{Underground Neutrino Detectors for Particle and Astroparticle Science: the 
Giant Liquid Argon Charge Imaging ExpeRiment (GLACIER)\footnote{To appear in the proceedings of DISCRETE'08: Symposium on Prospects in the Physics of Discrete Symmetries, Valencia, Spain, 11-16 Dec 2008.}
}

\author{Andr\'e Rubbia}

\address{ETH Zurich, Institute for Particle Physics, CH--8093 Z\"urich, Switzerland}

\ead{andre.rubbia@phys.ethz.ch}

\begin{abstract}
The current focus of the CERN program is the Large Hadron Collider (LHC), however, CERN is engaged in long baseline neutrino physics with the CNGS project and supports T2K as recognized CERN RE13, and for good reasons: a number of observed phenomena in high-energy physics and cosmology lack their resolution within the Standard Model of particle physics; these puzzles include the origin of neutrino masses, CP-violation in the leptonic sector, and baryon asymmetry of the Universe. They will only partially be addressed at LHC.
A positive measurement of $\sin^22\theta_{13}>0.01$ would certainly give a tremendous boost to neutrino physics by opening the possibility to study CP violation in the lepton sector and the determination of the neutrino mass hierarchy with upgraded conventional super-beams. These experiments (so called ``Phase II'') require, in addition to an upgraded beam power, next generation very massive neutrino detectors with excellent energy resolution and high detection efficiency in a wide neutrino energy range, to cover 1st and 2nd oscillation maxima, and excellent particle identification and $\pi^0$ background suppression.
Two generations of large water Cherenkov detectors at Kamioka (Kamiokande and Super-Kamiokande) have been extremely successful. And there are good reasons to consider a third generation water Cherenkov detector with an order of magnitude larger mass than Super-Kamiokande for both non-accelerator (proton decay, supernovae, ...) and accelerator-based physics.
On the other hand, a very massive underground liquid Argon detector of about 100 kton could represent a credible alternative for the precision measurements 
of ``Phase II'' and aim at significantly new results in neutrino astroparticle and non-accelerator-based particle physics (e.g. proton decay).
\end{abstract}

\section{Introduction}
The field of neutrino physics was born in 1930 when Wolfgang Pauli 
postulated the new particle to save conservation of energy, momentum and angular momentum
in beta decay. Today, Europe is engaged in long baseline neutrino physics with
the CNGS project~\cite{CNGS} and T2K~\cite{Itow:2001ee}, hoping to shed
some light on a number of observed phenomena in high-energy physics and cosmology lacking their resolution 
within the Standard Model of particle physics. These puzzles include the origin of neutrino masses, 
CP-violation in the leptonic sector, and baryon asymmetry of the Universe. 
They can only be partially addressed at LHC.

The CNGS has recently begun operation and first events have been collected~\cite{Acquafredda:2006ki}
 in OPERA~\cite{opera}. The goal 
is to unambiguously detect 
the appearance of $\tau$ leptons induced by $\nutau$ CC events,
thereby proving the $\numu\rightarrow\nutau$ flavor oscillation mechanism.
The OPERA result, together with well established observations of
solar and atmospheric neutrinos, in particular from Superkamiokande~\cite{Kajita:2006gs}, SNO~\cite{Ahmad:2002jz} 
and KamLAND~\cite{Eguchi:2002dm}, will most likely confirm the validity of the $3\times 3$ Pontecorvo-Maki-Nakagawa-Sakata (PMNS)~\cite{pontecorvo} mixing
matrix approach to describe all the observed neutrino flavor conversion phenomena. The PMNS 
is usually parameterized as:
\begin{eqnarray}
U & = & \left[ \begin{tabular}{ccc} $U_{e1}$ &  $U_{e2}$ &  $U_{e3}$ \\
$U_{\mu1}$ &  $U_{\mu2}$ &  $U_{\mu3}$ \\
$U_{\tau1}$ & $ U_{\tau2}$ &  $U_{\tau3}$ \end{tabular} \right] \nonumber \\
& = & 
\left[ \begin{tabular}{ccc} $1$ &  $0$ &  $0$ \\
$0$ &  $c_{23}$ &  $s_{23}$ \\
$0$ & $-s_{23}$ &  $c_{23}$ \end{tabular} \right]
\left[ \begin{tabular}{ccc} $c_{13}$ &  $0$ &  $s_{13}e^{-i\delta}$ \\
$0$ &  $1$ &  $0$ \\
$0$  & $-s_{13}e^{-i\delta}$ &  $0$ \end{tabular} \right]
\left[ \begin{tabular}{ccc} $c_{12}$ &  $s_{12}$ &  $0$ \\
$-s_{12}$ &  $c_{12}$ &  $0$ \\
$0$ & $0$ &  $1$ \end{tabular} \right]
\left[ \begin{tabular}{ccc} $e^{i\alpha/2}$ &  $0$ &  $0$ \\
$0$ &  $e^{i\alpha_2/2}$ &  $0$ \\
$0$ & $0$ &  $1$ \end{tabular} \right]\nonumber \\
& = & 
\left[ \begin{tabular}{ccc} $c_{12}c_{13}$ &  $s_{12}c_{13}$ &  $s_{13}e^{-i\delta}$ \\
$-s_{12}c_{23}-c_{12}s_{23}s_{13}e^{i\delta}$ &  $c_{12}c_{23}-s_{12}s_{23}s_{13}e^{i\delta}$ &  $s_{23}c_{13}$ \\
$s_{12}s_{23}-c_{12}c_{23}s_{13}e^{i\delta}$ &  $-c_{12}s_{23}-s_{12}c_{23}s_{13}e^{i\delta}$ &  $c_{23}c_{13}$ 
\end{tabular} \right]
\left[ \begin{tabular}{ccc} $e^{i\alpha/2}$ &  $0$ &  $0$ \\
$0$ &  $e^{i\alpha_2/2}$ &  $0$ \\
$0$ & $0$ &  $1$ \end{tabular} \right]
\end{eqnarray}
where $s_{12} = \sin\theta_{12}$, $c_{12} = \cos\theta_{12}$, etc. 
The phase factors $\alpha_1$ and $\alpha_2$ are non-zero only if neutrinos are Majorana particles 
(whether they are or not is at present unknown), but do not enter into flavor oscillation phenomena.
The Dirac phase $\delta$, relevant to neutrino flavor oscillations, is non-trivial only if the three mixing angles $\theta_{12}$, $\theta_{23}$, $\theta_{13}$ do not vanish.
In this case, the condition $\delta\ne 0$ 
induces different flavor transition probabilities for neutrinos and
antineutrinos. 
In order to complete the PMNS picture, all the elements (magnitude
and phase) of the matrix must be precisely determined. That includes the
$U_{e3}$ element for which today there is only an upper bound 
from the CHOOZ~\cite{Apollonio:1999ae} reactor experiment, giving
in the standard parameterization $\sin^22\theta_{13}\lesssim 0.1$ (90\%C.L.).

The determination of the unknown element of the PMNS matrix is possible via the study of
$\numu\rightarrow\nue$ oscillations at the baseline relevant for
atmospheric neutrinos.
T2K and NOvA~\cite{Ayres:2004js} under construction should reach
a sensitivity $\sin^22\theta_{13} \lesssim 0.01$ (90\%C.L.).
A very large fraction of the European neutrino community is engaged
in T2K, with participations from France, Germany, Italy, Poland, Spain, Switzerland and United Kingdom.
The T2K beam will start commissioning in April 2009 and the first physics results are
expected for the summer 2010.
DOUBLE-CHOOZ~\cite{Ardellier:2004ui} located in northern France will also look for
a small $\nue\rightarrow\nu_x$ disappearance effect from reactors.

After $\theta_{13}$, the measurement of $\delta$ is one of the main challenges of future
neutrino oscillation experiments. Due to matter effects, neutrinos and
antineutrinos propagate differently through the Earth. This will also induce
differences in oscillatory behaviors of neutrinos and antineutrinos that will affect the sensitivity of
an unambiguous determination of the value of the mixing matrix complex phase.

\section{Neutrino astrophysics}
Two generations of large water Cherenkov detectors at Kamioka (Kamiokande\cite{Koshiba:mw} and Super-Kamiokande\cite{Fukuda:2002uc})
have been very successful in research of neutrino physics with astrophysical sources from the Sun, from supernovae 
and atmospheric neutrinos.

Core collapse supernovae are a huge source of all flavor
neutrinos. 
The flavor composition, energy spectrum and time structure of
the neutrino burst  from a galactic supernova can provide information
about the explosion mechanism
and the mechanisms of proto neutron star cooling. 
Such data can also give information
about the intrinsic properties of the neutrino such as flavor
oscillations. 
One important question is to understand to which extend can the
supernova and the neutrino physics be decoupled in the observation
of a single supernova. On one hand, the understanding of the supernova
explosion mechanism is still plagued by uncertainties which have an impact
on the precision with which one can predict time, energy and flavor-dependent
neutrino fluxes. On the other hand, the neutrino mixing properties are
not fully known, since the type of mass hierarchy and the value of
the  $\theta_{13}$ angle are unknown, and in fact large uncertainty still exists on
the prediction of the actual effect of neutrino oscillations in the event
of a supernova explosion.
Future detection of a supernova neutrino burst by large underground detectors will give important 
information for the explosion mechanism of collapse-driven supernovae. 

It is also believed 
that the core-collapse supernova explosions have traced the star formation 
history in the universe and have emitted a great number of neutrinos, which 
should make a diffuse background~\cite{Ando:2002ky}. Detection of this supernova relic neutrino (SRN) background
is one of the goals of the currently working large neutrino detectors, 
Super-Kamiokande (SK) and Sudbury Neutrino Observatory (SNO).

Neutrinos are produced in abundance by high-energy cosmic rays impinging on 
the Earth's atmosphere. These neutrinos, spanning energies from a few MeV to 
the highest-energy cosmic rays, provide a background against which one must discriminate 
to detect extraterrestrial sources, but they have been used
to detect muon neutrino oscillations, providing the 
first convincing evidence for neutrino mass~\cite{Kajita:2006gs}.  

\section{A theoretical hint ?}
The neutrino roadmap (see e.g.~Ref.~\cite{Bandyopadhyay:2007kx}) depends strongly on the value 
of $\theta_{13}$:
a positive signal compatible with $\sin^22\theta_{13}\gtrsim0.01$ would give a tremendous boost 
to neutrino physics by opening the possibility to study CP violation in the lepton sector and
mass hierarchy determination
with upgraded conventional superbeams.
The community is therefore presently contemplating next generation
experiments (so called ``Phase II'') which require, in addition to upgraded beam power, 
next generation very massive neutrino detectors.
If $\sin^22\theta_{13}\lesssim0.01$, then most likely
new advanced neutrino beams, like for instance beta-beams or neutrino factories, will be
required.

\begin{figure}[h]
\includegraphics[width=20pc]{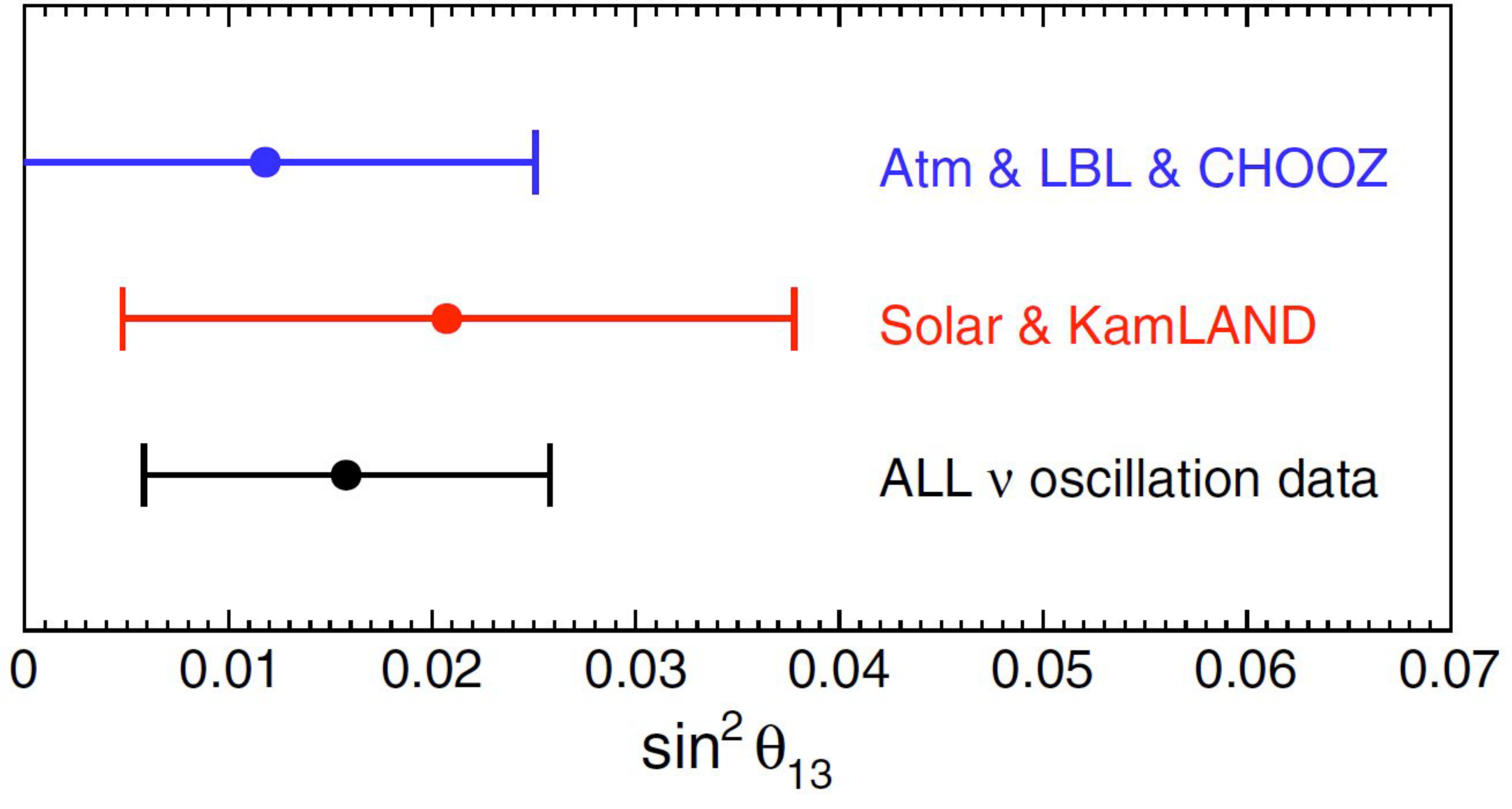}\hspace{2pc}%
\begin{minipage}[b]{14pc}\caption{\label{fig:fogliclaim}The global
neutrino oscillation analysis yielding a positive effect with allowed
$1\sigma$ ranges for $\sin^2\theta_{13}$ from different input data. From Ref.~\cite{Fogli:2008jx}.}
\end{minipage}
\end{figure}

Is there a hint?
A global analysis of alll neutrino oscillation data, focusing on $\theta_{13}$
finds  two converging hints of $\theta_{13}>0$, each at the level of $\simeq 1\sigma$~\cite{Fogli:2008jx}: an older one coming from atmospheric neutrino
data, and a newer one coming from the combination of solar and long-baseline reactor neutrino data. Their
combination provides the global estimate $\sin^2 \theta_{13} = 0.016\pm 0.010(1\sigma)$, implying a preference for  $\theta_{13}>0$ 
with non-negligible statistical significance ( $\simeq  90\% C.L.$). See Figure~\ref{fig:fogliclaim}.
Is this result due to low statistics?
T2K and NOvA will be key experiments to settle this question.

\section{Matter dominance, Baryogenesis and Leptogenesis, Baryon number violation}

The present Universe is known to be entirely dominated by matter over anti-matter. 
Following Sakharov~\cite{Sakharov:1967dj} Baryogenesis requires Baryon-number-violation and a matter-antimatter asymmetry in fundamental reactions among elementary particles.
In a related mechanism named leptogenesis, leptons could also have participated in the process. The study of CP-violation in the leptonic
sector might shed some light~\cite{Pascoli:2003uh}.

Baryon number violation implies that the proton is unstable.
In the simplest Grand Unified Theories (GUT)\cite{gut, Georgi:1974sy}, nucleon decay proceeds via 
an exchange of a massive boson $X$ between
two quarks in a proton or in a bound neutron.  In this reaction, one quark
transforms into a lepton and another into an anti-quark which binds with a
spectator quark creating a meson.  
According to the experimental results from
Super-Kamiokande~\cite{Shiozawa:1998si,Shiozawa:2003it} constraining
the partial decay to 
$\tau/B(p \rightarrow e^+ \pi^0) > 5.4 \times 10^{33}$~years (90\%C.L.), the minimal
SU(5)~\cite{Georgi:1974sy}, predicting a proton
lifetime proportional to $\alpha^{-2}M_X^4$ where $\alpha$
is the unified coupling constant and $M_X$ the mass of
the gauge boson $X$, seems definitely ruled out.
In addition, in this model it does not seem possible to achieve the unification of gauge couplings 
in agreement with the experimental values of the gauge couplings at the
$Z^0$ pole~\cite{Amaldi:1991cn}.

Supersymmetry, motivated by the so-called ``hierarchy problem'',
 postulates that for every SM particle, there is a
corresponding ``superpartner'' with spin differing by 1/2 unit from the SM
particle \cite{susy}. In this case,
the unification scale turns out higher, and pushes up the proton lifetime in the $p \rightarrow e^+ \pi^0$
channel up to $10^{36\pm1}$~years, compatible with experimental results.
At the same time, alternative decay channels open up via
dimension-five operator interactions with the exchange of heavy
supersymmetric particles.
In these models, transitions from one quark family in the initial state to the same
family in the final state are suppressed.  
Since the only second or third generation quark
which is kinematically allowed is the strange quark, an anti-strange quark
typically appears in the final state for these interactions. The anti-strange
quark binds with a spectator quark to form a $K$ meson in the final state~\cite{Pati}.
The searches
for decays $p \to \bar{\nu} K^+$, $n \to \bar{\nu} K^0$, $p \to \mu^+ K^0$ and $p \to e^+ K^0$ 
modes were also performed in Super-Kamiokande~\cite{Hayato:1999az,Kobayashi:2005pe}
yielding counts compatible with background expectations, leading to limits on possible
minimal SUSY SU(5) models~\cite{Dimopoulos:1981zb,Sakai:1982pk,Hisano:1993jj}.
The theoretical predictions, however, vary
widely, since there are many new unknown parameters
introduced in these models.
Other alternative models have been discussed in the litterature~\cite{Nath:1985ub, Nath:1998kg, Shafi:1999vm, Lucas:1997bc,
Pati:2003qi, Babu:1998js, Babu:1998wi, Pati:2000wu,Ellis:2002vk,Arkani-Hamed:2004yi,Dorsner:2005fq,Hebecker:2002rc,
Alciati:2005ur, Klebanov:2003my}. 
In addition to the above mentioned GUTs, other supersymmetric SUSY-GUT, SUGRA unified models, unification based on extra dimensions, and string-M-theory models
 are also possible (see Ref.~\cite{Nath:2006ut} for a review).  
 All these models predict nucleon instability at some level.
It is also worth noting that 
 theories without low-energy super-symmetry~\cite{Arkani-Hamed:2004yi,Sayre:2006ma,Wiesenfeldt:2006ut}
predict nucleon decay lifetimes in the range $10^{35\pm1}$~years.
  Finally, we point out Ref.~\cite{Dorsner:2004xa} where an upper bound on
 nucleon lifetime is derived.

\section{Frontier physics requires next generation very large underground detectors}

The community is addressing a class of ``ultimate'' generation
experiments for the search of CP-violation in neutrino oscillations and
the determination of the neutrino mass hierarchy. 
New generation superbeams or beta-beams need giant detectors.
Neutrino factories require large magnetized detectors. These large
underground detectors would also advance the field of
neutrino astrophysics and look for proton decays in the range of $10^{34}-10^{35}$~yr.
Experimental aspects of
nucleon decay detection were discussed in Ref.~\cite{Rubbia:2004yq}.

\subsection{Requirements on long baseline neutrino detectors}
T2K and NOvA as ``Phase I'' experiments,  will improve
the sensitivity on $\theta_{13}$  by about an order of magnitude compared
to CHOOZ. In addition, the 
NOvA experiment, due to its longer baseline (810 km compared to 295 km of T2K), would have the 
ability to determine the neutrino mass hierarchy if $\theta_{13}$ is close to the current CHOOZ limit. 
We note that  ``Phase I'' experiments do not have significant discovery potential for CP violation. 
Beyond this step,  the goals of ``Phase II'' experiments are (to some extent in order of priority): 
\begin{enumerate}
\item To have a discovery potential for measuring CP violation in the neutrino sector, in
case of $\theta_{13}$ discovery in ``Phase I'' experiments. 
\item To extend the discovery potential for determining the neutrino mass hierarchy for, at least, 
the region of the $\theta_{13}$  discovery potential of ``Phase I'' experiments. 
\item To extend the $\theta_{13}$ discovery potential, in case ÓPhase IÓ experiments have only 
yielded more stringent limits. 
\end{enumerate}

A ``Phase II'' long-baseline neutrino oscillation experiment is necessarily ``ultimate''
since it should~\cite{Rubbia:2004tz}:
\begin{enumerate}
\item be designed to have ample statistics to precisely
determine the oscillation probability as a function of the neutrino energy;
\item have an excellent energy resolution to observe the energy dependence of
the oscillation probability and help lift degeneracy of the parameters governing
the neutrino oscillations (see e.g. Ref.\cite{Bueno:2001jd});
\item be performed at a wide-band neutrino beam to cover enough ``oscillations'' peaks
or do ``counting'' at different neutrino beam energy settings;
\item have the possibility to study neutrinos and antineutrinos ideally separately
in order to lift degeneracies (even in the counting mode).
\end{enumerate}

\begin{figure}[h]
\includegraphics[width=25pc]{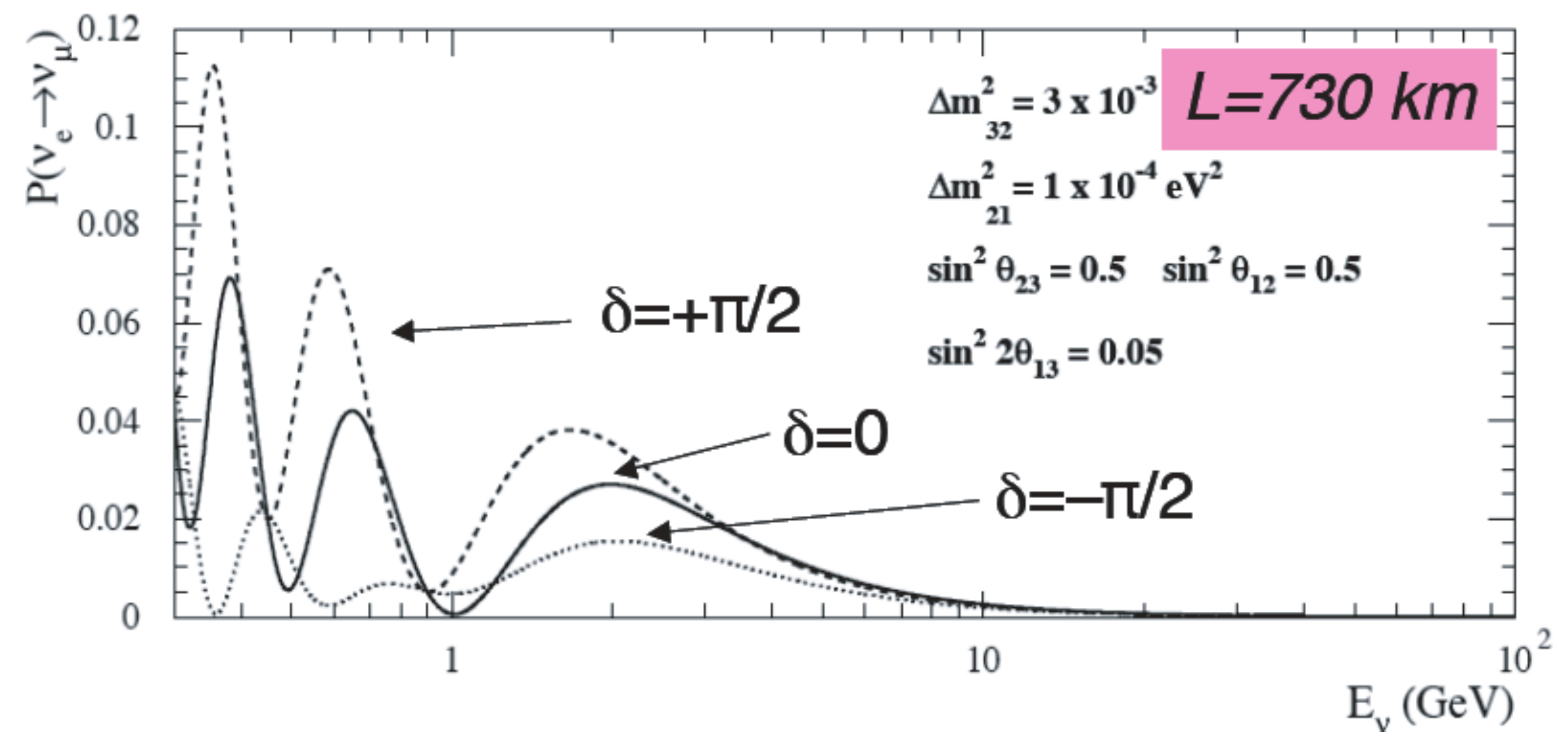}\hspace{2pc}%
\begin{minipage}[b]{9pc}\caption{\label{fig:oscillgraph}The $\numu\rightarrow\nue$ oscillation probability as a function of neutrino
energy for different values of the $\delta$ phase for a given set of oscillation 
parameters and a distance of 730 km. From Ref.~\cite{Rubbia:2004tz}.}
\end{minipage}
\end{figure}

Figure~\ref{fig:oscillgraph} illustrates qualitatively the fact that a measurement of the oscillation
probability {\it as a function of energy with good resolution} would indeed
provide direct information on the $\delta$-phase, since this latter introduces
a well-defined energy dependence of the oscillation probability, which is different
from the, say, energy dependence introduced by $\theta_{13}$ alone (when $\delta=0$).

The two detector technologies considered are a massive deep underground Water
Cherenkov imaging (WC) detector with a fiducial mass of 300-500 kton, and a fully
active finely grained liquid Argon time-projection-chamber (LAr TPC)
with a mass of $\sim$ 100 kton, as described in the following sections.

\subsection{Large Water Cherenkov Imaging detector}
Super-Kamiokande is composed of a tank of 50~kton of water (22.5~kton fiducial) which is
surrounded by 11146 20-inch phototubes immersed in the water. About $170~\gamma/cm$ are produced
by relativistic particles in water in the visible wavelength $350<\lambda<500~nm$. With 40\% PMT coverage
and a quantum efficiency of 20\%, this yields $\approx 14$ photoelectrons per cm or $\approx 7$~p.e. per MeV 
deposited.

There are good reasons to consider a third generation water Cherenkov detector with an
order of magnitude larger mass than Super-Kamiokande: a megaton Water Cherenkov detector will have a 
broad physics programme, including both
non-accelerator (proton decay, supernovae, ...) and accelerator physics.

Hyper-Kamiokande~\cite{Nakamura:2003hk} has been
proposed with about 1~Mton, or about 20 times as large as Super-Kamiokande, based
on a trade-off between physics reach and construction cost. 
Further scaling is limited by light propagation in water (scattering, absorption).
Although this order of magnitude
extrapolation in mass is often considered as straight-forward, a number of R\&D efforts
are needed. 
An important item for Hyper-Kamiokande is developments of new photo-detectors: with the
same photo-sensitive coverage as that of Super-Kamiokande, 
the total number of PMTs needed for Hyper-Kamiokande will be $\simeq 200'000$. Possibilities
to have devices with higher quantum efficiency, better performance, and cheaper
cost are being pursued. 

\subsection{Large liquid argon Time Projection Chamber (TPC)}

Among the many ideas developed around the use of liquid noble gases, the Liquid 
Argon Time Projection Chamber (LAr TPC) (See Refs.~\cite{icarusprops,t3000,Amerio:2004ze}
and references therein)  certainly represented one of the most
challenging and appealing designs.
The LAr TPC is a powerful detector for homogeneous and high accuracy imaging of potentially very massive active volumes. 
It is based on the fact that in highly pure Argon, ionization tracks can be drifted
over distances of the order of meters. 
Imaging is provided by position-segmented electrodes at the end of the drift path, continuously recording the 
signals induced. $T_0$ is provided by the prompt scintillation light. 

Images taken with a liquid Argon TPC are comparable with pictures from bubble 
chambers. As it is the case in bubble chambers, events can be analyzed by reconstructing 
3D-tracks and particle types for each track in the event image, with a lower energy 
threshold of few MeV for electrons and few tens of MeV for protons. The particle type 
can be determined from measuring the energy loss along the track ($dE/dx$) or from 
topology (i.e. observing the decay products). Additionally, the electronic readout allows
to consider the volume as a calorimeter adding up all the collected ionization charge.
The calorimetric performance can be excellent, depending on event energy and topology.
Liquid Argon detectors have therefore advantages on
\begin{itemize}
\item good energy resolution/reconstruction,
\item good background suppression,
\item good signal efficiency.
\end{itemize}

The liquid Argon TPC imaging should offer optimal conditions
to reconstruct the electron appearance signal in the energy
region of interest in the GeV range, while considerably 
suppressing
the NC background consisting of misidentified $\pi^0$'s. 
The signal efficiency
is expected to be higher to that of the Water Cerenkov detectors, hence, the LAr TPC detector
could be smaller for similar performance. In addition, a LAr TPC should allow operation at shallow
depth and the
constraints on the excavation and the related siting issues of the detector
should hence be reduced.
Nonetheless, {\bf in order to compete with next generation megaton-scale Water Cerenkov detectors,  a liquid Argon detector 
should have a mass at the level of 100~kton~\cite{Rubbia:2004tz}}.
Since 2004, this result has being extensively discussed and supported in the 
literature~\cite{Schwetz:2008tc,Huber:2008yx, Badertscher:2008bp, Meregaglia:2008qr, Barger:2007jq, Meregaglia:2006du, Mena:2005ri, Cline:2005dm}.
This conclusion also holds for non-accelerator proton decay searches~\cite{Bueno:2007um} and
detection of astrophysical neutrino sources~\cite{Autiero:2007zj,Barger:2005it, Cocco:2004ac, GilBotella:2004bv}.

\section{Discovery potential of very large LAr TPCs at CNGS}

In 2008 the CNGS integrated intensity delivered to target was $1.8\times 10^{19}$~pots.
The current CNGS optimization provides limited
sensitivity to the $\nu_\mu\rightarrow\nu_e$ reaction and OPERA should ultimately reach
a sensitivity $\sin^22\theta_{13} \lesssim 0.06$ (90\%C.L.) in 5 years of
running with the nominal $4.5\times 10^{19}$~pot/yr \cite{Komatsu:2002sz}.
The ICARUS~T600~\cite{Amerio:2004ze}, still to be
commissioned, will detect too few contained CNGS events to competitively 
study electron appearance. Ways to improve the $\theta_{13}$-sensitivity
at CNGS have been discussed in the past~\cite{Ball:2006uw,Rubbia:2002rb}.

\begin{table}[tb]
  \caption{Design parameters for the various beams at J-PARC. Comparison with the 
  dedicated CNGS intensity and assumed upgrades of the CERN SpS 
  complex (see text). The beam power corresponds to the instantaneous power
  on the neutrino target while the product $E_p\times N_{pot}$ corresponds
  to the total amount of energy deposited on the target per year, which is more
  relevant to calculate neutrino event rates. The nominal CNGS year is assumed
  to deliver $4.5\times 10^{19}$~pots.}
  \label{tab:accelerators}
   \begin{center}
    \begin{tabular}{|l|c|c|c|c|c|c|c|}
\br
      & \multicolumn{3}{c|}{J-PARC} &  \multicolumn{4}{c|}{CERN SpS}\\   
       &   design & upgrade  & ultimate            &     CNGS  & + & 1& 2\\
       &        \cite{Itow:2001ee}           &  \cite{NP08}         & 
       \cite{Itow:2001ee}Ê& dedicated & \cite{Meregaglia:2006du} &\cite{Baibussinov:2007ea} & \cite{Baibussinov:2007ea} \\
              \mr
              Proton energy $E_p$ & \multicolumn{2}{c|}{30 GeV} & 40 GeV &  \multicolumn{4}{c|}{400 GeV}\\
               $ppp (\times 10^{13})$ & 33 & 67 & $>67$ & 4.8 & 14& 4.8 & 15\\
               $T_c$ (s) & 3.64 & 2 &  $<2 $ & 6 & 6 & 6&6\\
               Efficiency &  1.0 & 1.0 & 1.0 & 0.55 &0.83& 0.8& 0.8\\
               Running (d/y) & 130  & 130 & 130 & 220& 220 & 240 & 280 \\
               $\bf N_{pot}$ $/$ \bf yr ($\bf \times 10^{19}$) &  \bf 100 & \bf 380 & $\bf\simeq 700$ & \bf 7.6 & \bf 33 & \bf 12 & \bf 43.3 \\
               Beam power (MW) & 0.6 & 1.6 & 4  &    0.5 & 1.5 & 0.5 & 1.6\\
               $E_p\times N_{pot}$  & 4 &11.5 & 28 & 3 & 13.2 & 4.7& 17.3\\
               ($\times 10^{22}$ GeV$\cdot $pot/yr) & & & & & & &  \\
              Relative increase & & $\times 3$ & $\times 7$ & $\times 2$ & $\times 7$ & $\times 3$ & $\times 10$ \\
               Timescale & $>2009$ & 2014? & $>$2014? & $>2008$ &  \multicolumn{3}{c|}{$>$2016 ?}  \\
               \br
      \end{tabular}
  \end{center}
  \end{table}

In 2006, the physics potential of an {\bf intensity upgraded} and 
energy re-optimized CNGS neutrino beam coupled to a 100~kton
liquid Argon TPC located at an appropriately chosen off-axis position was
studied~\cite{Meregaglia:2006du}. The study concluded that improvements in $\theta_{13}$
reach,  sensitive searches for CP-violation and mass hierarchy
determination were possible and potentially competitive with J-PARC~\cite{NP08}.
The discussion relied on the observation that whereas J-PARC provides
a rapid cycle with high intensity proton bunches at $\sim 40$~GeV/c,
the CERN proton complex has fewer protons and a slower 
cycle but can accelerate up to 400~GeV/c.
Hence, the  resulting target beam powers are -- on paper -- comparable
(See Table~\ref{tab:accelerators}). 
In particular, it was noted
that future upgrades of the CERN LHC injection chain
could provide increased proton intensities in
the SPS. This option labelled ``CNGS+" in Table~\ref{tab:accelerators}
accordingly envisioned $3.3 \times 10^{20}$ pots/yr.

The same idea was subsequently proposed assuming
a smaller detector of 20~kton 
located at an angle OA~0.8$^o$ at a baseline of 730~km 
 (MODULAr~\cite{Baibussinov:2007ea}).
In this case, two possible upgrades for CNGS beam labelled as ``CNGS1" and ``CNGS2"
yielding $1.2 \times 10^{20}$ pot/yr and $4.33 \times 10^{20}$ pot/yr
were considered.

\begin{figure} [tbh]
\begin{center}
     \includegraphics[width=.95\textwidth]{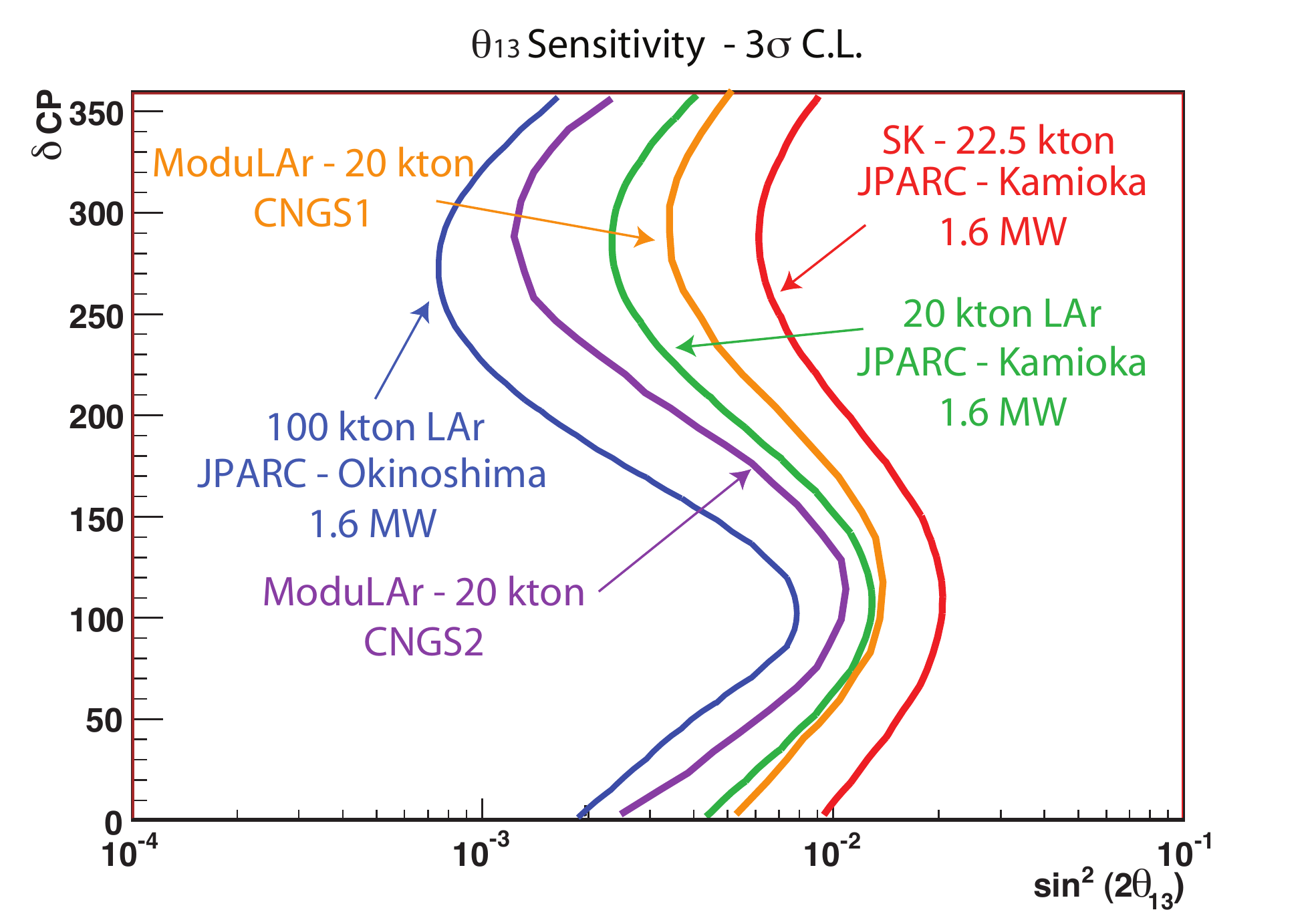}
\caption{$\theta_{13}$ sensitivity at $3\sigma$~C.L. for  20~kton LAr detector at 295~km, 2.5 degrees off-axis for 5 years of neutrino beam at 1.6~MW (green line). For comparison the sensitivity of T2K (22.5~kton WC at Kamioka - 1.6~MW) and GLACIER-100~kton at Okinoshima, and the sensitivity of  MODULAr-20~kton experiment~\cite{Baibussinov:2007ea} for two different upgrades of CNGS beam are shown (see text for details).
{\bf The J-PARC 1.6~MW represents an increase $\times 3$ relative to design intensity while CNGS2 represents $\times 10$ compared to the CNGS design}.}
\label{fig:ALL1}
\end{center}
\end{figure}

In order to quantitatively compare potential CNGS upgrades with possible options at J-PARC, we
focus on the $\sin^22\theta_{13}$ sensitivity. The obtained conclusions can
be readily extrapolated to the CP-violation and mass hierarchy sensitivity reach.
Figure~\ref{fig:ALL1} shows the expected $\theta_{13}$ sensitivity at
the $3\sigma$ C.L. for a 20~kton LAr TPC detector located at Kamioka
after 5 years of neutrino run with an upgraded J-PARC beam power of 1.6~MW~\cite{NP08}.
The expected sensitivity with the
existing Super-Kamiokande detector and 1.6~MW is also shown. A 20~kton LAr TPC at
Kamioka is an effective way to improve the $\theta_{13}$-sensitivity
of the T2K experiment.  The MODULAr expectation~\cite{Baibussinov:2007ea} is shown
for comparison. Finally, the sensitivity
of a 100~kton LAr detector at Okinoshima~\cite{Badertscher:2008bp} is plotted. 
We note that J-PARC 1.6~MW represents an increase $\times 3$ relative to design 
intensity while CNGS2 represents $\times 10$ compared to the CNGS design.

A CERN report~\cite{Meddahi:2007ju} eventually indicated that
with an upgrade of the SPS RF and new injectors, it would be possible
to accelerate $2.4 \times 10^{20}$ pot/yr at the SPS (timescale $>$2016). This means that the CNGS+ 
exposure of Ref.~\cite{Meregaglia:2006du} would correspond to a run of 7 years instead of 5 years 
and the CNGS2 beam of Ref.~\cite{Baibussinov:2007ea} to 9 years instead of the assumed 5 years.
Yet intensity limitations will not only come from the accelerator complex, but also from the design of the equipment
in the current CNGS facility and from radiation and waste issues.

In conclusion, calculations show that the $\sin^22\theta_{13}$ reach and the searches for CP-violation and mass 
hierarchy are competitive with future options at J-PARC if the CNGS beam intensity can be increased compared to its design 
value $4.5\times 10^{19}$~pot/yr by a factor $\times 3-\times 10$.
Yet CNGS intensity limitations do not only come from the performance of the accelerator complex. 
An upgraded CNGS -- competitive with JPARC -- will require a re-classification and/or partial reconstruction 
of the neutrino beam-line infrastructure, raising questions of feasibility, timescale and costs.

\section{Potential European underground sites for very large underground detectors}

\subsection{The FP7 design studies}
In the spirit of an European-wide coordination,
potential future neutrino projects in Europe beyond the CNGS are presently being studied in two design studies supported by the EC FP7 programme: 
\begin{enumerate}
\item the {\bf EuroNU} design study dedicated to the assessment and technical development of a low energy superbeam and of next generation high-intensity beta-beams and neutrino factories, and 
\item the  {\bf LAGUNA} design study dedicated to the feasibility of very large underground infrastructures able to host next generation neutrino physics and astroparticle physics and proton decay experiments. 
\end{enumerate}
Although the studies are tailored to Europe and are discussed within the context of a future European programme 
involving CERN, the results of these studies are often applicable to a general world-wide context, 
in particular, many concepts and R\&D results could be applied also at future neutrino 
experiments in Japan involving JPARC, or elsewhere. 

\subsection{Seven potential sites in Europe}
The LAGUNA design study considers three different detector technologies~\cite{Autiero:2007zj}, and seven potential underground sites (see Table~\ref{tab:lagunasite}) in order to identify the 
scientifically and technically most appropriate and cost-effective strategy for future large-scale underground detectors in Europe.
The main deliverable will be a report, which should contain relevant information for a forward decision, which is due in Fall 2010.

One of the seven LAGUNA site is located at an off-axis angle $\simeq 1.0^o$ of the CNGS beam~\cite{Meregaglia:2006du}.
In addition, given the intensity limitation of the CNGS design, the option of a completely new high-intensity 
neutrino beam line from CERN towards one of the LAGUNA sites is envisaged. The baselines
from CERN span a large variety of possibilities, ranging from the shortest distance at Fr\'ejus of 130~km, 
with Canfranc and Caso at about 650~km, Polkowice and Boulby at about 1000~km, Slanic at
about 1600~km and finally Pyhasalmi at the longest 2300~km. Table~\ref{tab:lagunasite}  lists
for comparison the neutrino energy corresponding to the first oscillation maximum (for $\Delta m^2=2.5\times 10^{-3}$~eV$^2$).

\begin{table}[h]
\caption{\label{tab:lagunasite}Seven potential sites for a new European underground laboratory for next generation
neutrino astrophysics and proton decay studied within the LAGUNA design study.}
\begin{center}
\begin{tabular}{|l|c|c|c|c|}
\br
Name & Type & Envisaged Depth&Distance from CERN &Energy 1st \\
& & (m.w.e) & (km) & osc. max (GeV) \\
\mr
Fr\'ejus (F) & Road tunnel & $\simeq 4800$ & 130 & 0.26 \\
Canfranc (ES) & Road tunnel & $\simeq 2100 $& 630 & 1.27  \\
Caso (IT) & Green field & $\simeq 1500$ & 665 ($\simeq 1.0^o$OA) & 1.34  \\
Polkowice (PL) & Mine & $\simeq 2400$  & 950&  1.92 \\
Boulby (UK) & Mine & $\simeq 2800$  & 1050 & 2.12 \\
Slanic (RO) & Salt mine & $\simeq 600$  & 1570 & 3.18  \\
Pyhasalmi (FI) & Mine & $\simeq 4000$  & 2300 & 4.65  \\
\br
\end{tabular}
\end{center}
\end{table}

The rock overburden is a relevant parameter for low background experiments to reduce
cosmogenic backgrounds in order to perform sensitive neutrino astrophysics and
proton decay searches, in addition to the beam related physics. From the three detector technologies
studied in LAGUNA~\cite{Autiero:2007zj}, the liquid Argon TPC option is the only one
that could be considered at rather shallow depths. The computed average number of muons entering the detector per unit time
for various geographical configurations is listed in Table~\ref{tab:nmninLar01}. The effective mass corresponds to the mass
of Argon that can be used when, in both 2D readout views,
 a slice of size 10~cm around each crossing muon is vetoed. From this table and from
 dedicated studies (see e.g. Ref~\cite{Bueno:2007um} for background in proton decay), it was concluded
 that large liquid Argon detector can fulfill all requirement at depths $\gtrsim 600$~m.w.e.

\begin{table}
\caption{Computed average number of muons entering the detector per unit time
for various geographical configurations. The effective mass corresponds to the mass
of Argon that can be used when, in both 2D readout views,
 a slice of size 10~cm around each crossing muon is vetoed. From Ref.~\cite{Bueno:2007um}.}
\label{tab:nmninLar01}
\centering
\begin{tabular}{|l|c|c|c|c|}\hline
\multicolumn{2}{|c|}{Depth} & \multicolumn{2}{|c|}{$E_\mu>1\ \mathrm{GeV}$} & Effective mass \\
Water equiv. & Standard rock &  Particles/s & Particles/10\,ms &  \\ \hline
\multicolumn{2}{|l|}{Surface detector} & 1300000& 13000 & -- \\ \hline
$\simeq 0.13$ km w.e. & $50\ \mathrm{m}$ &  10000& 100 & 50~kton\\ \hline
$\simeq0.5$ km w.e.& $188\ \mathrm{m}$ & $320$& 3.2 & 98~kton\\ \hline
$\simeq1$ km w.e.& $377\ \mathrm{m}$ &  $65$& 0.65 & 100~kton\\ \hline
$\simeq2$ km w.e.& $755\ \mathrm{m}$  & 6.2& 0.062& 100~kton \\ \hline
$\simeq3$ km w.e.& $1.13\ \mathrm{km}$ & $0.96$&0.01& 100~kton\\ \hline
\hline
\end{tabular}
\end{table}

\begin{figure}[h]
\includegraphics[width=25pc]{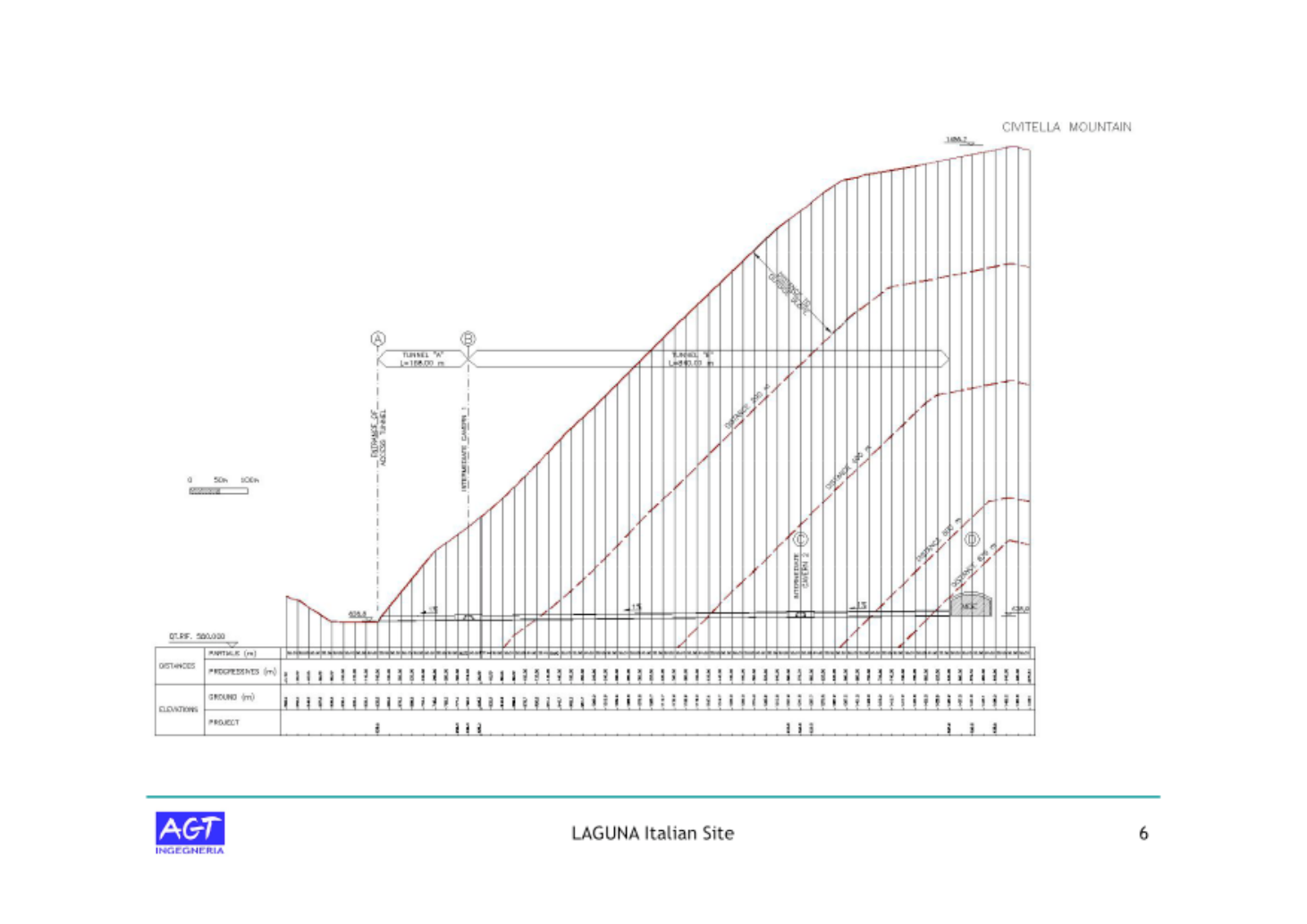}\hspace{2pc}%
\begin{minipage}[b]{10pc}\caption{\label{fig:agtshallow}Mountain profile for a shallow depth ($\simeq 1500$~m.w.e) 
being studied within the LAGUNA design study in collaboration with AGT Ingegneria~\protect\cite{AGT}.}
\end{minipage}
\end{figure}

The Caso site was chosen in collaboration with AGT Ingegneria~\cite{AGT} Êafter an investigation under the constraint of the present CNGS beam profile.
It was found by analyzing geological data and spotting a place in which
a horizontal road tunnel  of appropriate length could be dug into a mountain, so as to reach sufficient
rock overburden. Figure~\ref{fig:agtshallow} shows the mountain profile of the chosen
Caso site. This configuration allows in principle maximum freedom in site selection.

In the LAGUNA context, the possibility to rely on an alternative source of medium-energy high-intensity protons from 
CERN (e.g. from a newly built PS2) to create a proper high intensity low energy neutrino
beam is under discussion. 
For the longer term and in absence of a positive result from T2K/NOvA/DOUBLE-CHOOZ/..., large underground detectors in the LAGUNA sites could be operated as well in 
connection with other, more advanced neutrino beams like for instance beta-beams or neutrino factories. Such beams are currently being studied within the FP7 design study EuroNU.

\section{Beyond conventional superbeams}

\subsection{Neutrino Betabeams}
The physics potential of the large LAr detector in a 
Betabeam~\cite{Zucchelli:2002sa} was first presented in~\cite{cpbetabeam}.
The imaging of the events and the high energy resolution in
the LAr TPC make the study of Betabeams very attractive.
The possibility to have separately pure $\nu_e$ and $\bar\nu_e$ beams
combined with a massive 100~kton detector would be an ideal
configuration to study the neutrino oscillation parameters, in particular
the $CP$-phase. 

We tentatively considered the following baselines: $L=130$~km, 400~km and 950~km
with their corresponding neutrino energies~\cite{cpbetabeam}. The energy of
the neutrinos is determined by the ion $\gamma$ factor in the storage ring.
One finds that the Betabeam optimization requires 
the longest possible baseline, as long as matter effects are small, in order to benefit
from (1) the rise of the neutrino cross-section (this is particularly true for 
antineutrinos) and (2) the reduction of momentum smearing introduced
by the Fermi motion. While the baselines
of 130 and 400~km would be nicely accommodated by the CERN SPS maximal energy, the optimal
energy of the 950~km baseline will require a machine able to accelerate
protons up to $\sim$ 1 TeV.


\subsection{Neutrino Factories}
The physics potential of a magnetized large LAr detector combined with a 
Neutrino Factory~\cite{Geer:1997iz} has been studied in~\cite{Bueno:2001jd,Campanelli:zq,
Rubbia:2001pk,Bueno:2000fg,Bueno:2000jy,Campanelli:we,Campanelli:wi,Bueno:1999yx,Bueno:1999wb,Campanelli:1999ez}.
General ideas for intermediate and long baseline scenarios have also been discussed in~\cite{Bueno:1998xy,Bueno:1998pj}.

In \cite{Bueno:2000fg} it was concluded that in order to fully address the oscillation processes at a Neutrino
Factory, the ideal detector should be capable of identifying  and measuring all three charged lepton flavors
produced in charged current interactions and of measuring 
their charges to discriminate the incoming neutrino helicity. 
Embedding the volume of Argon inside a magnetic field would not
alter the imaging properties of the detector and
the measurement of the bending of charged hadrons or penetrating
muons would allow a precise measurement of the momentum and 
a determination of their charge.

For long penetrating tracks like muons, a field of $0.1$~T allows
to discriminate with $>3\sigma$ the charge for tracks longer than 4~m. This
corresponds to a muon momentum threshold of 800~MeV.
Hence, performance is excellent even at very low momenta.
Unlike muons or hadrons, the early showering of electrons 
makes their charge identification difficult. The track length
usable for charge discrimination is limited to a few radiation
lengths after which the shower makes the recognition of
the parent electron more difficult. 
From full simulations one found that the determination
of the charge of electrons in the energy range between
1 and 5~GeV is feasible with good purity, provided the field has a strength in the range of 1~T.
Preliminary estimates show that these electrons exhibit an average curvature 
sufficient to have electron charge discrimination better than
1$\%$ with an efficiency of 20\%.

From quantitative analyses of neutrino oscillation scenarios one can conclude~\cite{Bueno:2000fg} 
that in many cases the discovery sensitivities and the measurements of
the oscillation parameters
are dominated by the ability to measure the muon charge.
However, there are cases where identification of electron and tau samples 
significantly contribute. Kinematical searches for $\tau$ appearance in the
context of very long baseline experiments have been discussed in \cite{Campanelli:we}. One could
then make a unique measurement of the $\nue\rightarrow\nu_\tau$ transition
performing a stringent unitarity test of the lepton mixing matrix.

Apart from being able to measure very precisely the magnitude of all the
elements of the mixing matrix, the more challenging and most interesting
goal of the Neutrino Factory will be the search for effects related to  $CP$-violation (however affected by
matter effects), and between time-reversed transitions (the so called $T$-violation
unaffected by matter effects). 

As shown in \cite{Bueno:2001jd}, the ability to
measure electron and muon charges is the only way to address
$T$-violation, since it implies the comparison between the appearance of $\nu_\mu$ ($\bar\nu_\mu$)
and $\bar\nu_e$ ($\nu_e$) in a beam of stored $\mu^+$ ($\mu^-$) decays as a function of the
neutrino energy. A magnetized LAr detector would
be unique in this respect.

\section{Astrophysical neutrinos with very large liquid Argon TPCs}
The astrophysical neutrino physics programme is naturally very rich for a 100~kton LAr observatory.
One expects about 10000 atmospheric neutrino events per year and about 100 $\nu_\tau$
charged current events per year from $\nu_\mu$ oscillations. These events are characterized by
the excellent imaging capabilities intrinsic to the LAr TPC and will provide an unbiased sample
of atmospheric neutrinos with an unprecedented quality and resolution. This will allow
for improved measurements of atmospheric events, compared to existing or planned studies
based on Cerenkov ring detection.

Solar neutrinos provide
about 324000 events per year with electron recoil energy above $\sim$5~MeV, whose energy
will be measured with high accuracy (the threshold depends on the actual radioactive
background conditions at the underground site). This will provide
the possibility to make precision measurements of the solar neutrino flux and to
study possible short and long term variations, for example,
related to the solar cycles.

The physics that can be performed via the observation of a core collapse
supernova has been discussed in~\cite{GilBotella:2004bv,Gil-Botella:2003sz,Bueno:2003ei}.
A galactic SN-II explosion at 10~kpc yields about 20000 events. Sensitivity to extragalactic
supernovae ($e.g.$ in Andromeda) should be possible. Relic SN neutrino fluxes can also
be addressed~\cite{Cocco:2004ac}. A characteristic feature of the LAr TPC is the accessibility to
several independent detection channels (elastic scattering off electrons, charged neutrino
and antineutrino, and neutral currents on Argon nuclei) which have different sensitivities
to electron-neutrino, anti-electron-neutrino and other neutrino flavors 
(muon and tau (anti)neutrinos). The study of all neutrino flavors from supernova explosion
would be performed in great detail by a LAr detector, in an appreciably 
better way when compared to water
Cerenkov detectors, which are mainly focusing on the $\bar\nu_e$ flavor. A high
sensitivity to $\nu_e$s is fundamental to study the shock breakout, namely, the neutrino burst from the core
collapse, preceding the cooling phase~\cite{Gil-Botella:2003sz}. In addition, the sensivitiy
to all flavors during the cooling phase allows to over-constrain 
the supernova and the flavor mixing parameters and, to some 
extent, disentangle neutrino from supernova physics~\cite{GilBotella:2004bv}.

\section{Matter stability and proton decay with very large liquid Argon TPCs}
The physics of the nucleon decay has been addressed in Ref.~\cite{Bueno:2007um}.
Direct evidence for GUT and baryon number violation represents one of the outstanding goals of particle
physics. Nucleon decay searches require a very good knowledge of the backgrounds
induced by atmospheric neutrinos. Precise understanding of the neutrino physics is therefore
a fundamental component for ultimate proton decay experiments.
A target of 100~kton = 6 $\times$ 10$^{34}$ nucleons yields a sensitivity
for protons of $\tau_p/Br > 10^{34}$ years $\times$ T(yr)$\times\ \epsilon$ at the 90\% C.L.
in the absence of background. This means that lifetimes in the range of $10^{35}$ years can be reached within 10~years
of operation. Channels like $p\rightarrow \nu K$ have been shown to be indeed essentially background
free, even at shallow depths~\cite{Bueno:2007um}. 



\section{The GLACIER detector design}
The new concept GLACIER~\cite{Rubbia:2004tz}, 
scalable to a single detector unit of mass 100 kton, was proposed in 2003: 
it relies on a cryogenic storage tank developed by the petrochemical industry 
(LNG technology) and on a novel method of operation called the LAr LEM-TPC. 
Its parameters are summarized in Table~\ref{tab:sumpar}.
The main features of the GLACIER design 
are discussed next.

\begin{table}[htb]
\caption{\label{tab:sumpar}
Summary parameters of the 100 kton GLACIER detector. From Ref.~\cite{Rubbia:2004tz}.
} 
\centering
\begin{tabular}{|p{0.35\linewidth}|p{0.55\linewidth}|}
\br
\small Dewar&$\Phi\approx$ 70~m, height $\approx 20~m$, passive perlite insulated, 
heat input $\approx 5$~W/m$^2$ \\
Argon Storage & Boiling argon, low pressure ($<$~100~mb overpressure)\\
Argon total volume & 73118 m$^3$ (height = 19 m),ratio 
area/volume$\approx$15\%\\
Argon total mass&{\bf 102365 TONS} \\
Hydrostatic pressure at bottom&$\approx 3$~atm\\
Inner detector dimensions & Disc $\Phi \approx$ 70~m located in gas 
   phase above liquid phase\\
Electron drift in liquid &20~m maximum drift, HV= 2MV for E=1~kV/cm,
v$_d\approx 2 $~mm/$\mu$s, max drift time $\approx$~10~ms \\
Charge readout views& LEM-TPC; 2 independent perpendicular views, 3~mm pitch, 
with charge amplification\\
Scintillation light readout & Yes (trigger), 1000 immersed WLS-coated
8''PMT\\
Visible light readout (option) & Yes (Cerenkov light), 27000 immersed 8''PMTs or
20\% coverage, single photon counting capability\\
\br
\end{tabular}
\end{table}

\subsection{ Single module cryo-tank based on industrial LNG technology }

Liquefied Natural Gas (LNG, $\geq 95\%$ CH$_4$) is used when volume is an issue, in particular, for storage.
The technical problems associated to the design of large cryogenic tanks,
their design, construction and safe operation have already been extensively addressed and 
solved by the petrochemical industry over several decades.
Many large LNG tanks are in service,  about 300 worldwide in 2003, and there is an increased 
demand on LNG as alternative source of energy. The LNG tank volumes vary from 70'000 to 200'000~$\mathrm m^3$
with erection times from 2 to 5 years. 
The tanks, classified according to their containment type: single, double or full containment,
 are defined by international design codes and standards (BS7777, EN1473, API std 620).
Hundreds of large LNG tanker ships transporting volumes up to 145'000~$\mathrm m^3$ 
cross the oceans every years.

The LNG industry and in particular LNG tanks have an excellent safety record.  
During the last 60 years there have been only two spontaneous ruptures of large refrigerated
tanks (in 1944 and 1977). In the first, the cause was attributed to brittle fracture
due to the steel used and the second was due to a failure of a weld that had been
repaired following a leak the previous year.
Nowadays severe leaks of LNG are simply discounted as a mode of failure.
Internal leaks and vapor release are more frequent and often correlated to refilling procedures
with new LNG. Another source of accidents are seismic events.
This excellent safety record is a 
result of several factors.  First, the industry has technically and operationally 
evolved to ensure safe and secure operations.  Technical and operational advances 
include everything from the engineering that underlies LNG facilities to operational 
procedures to technical competence.  Second, the standards, codes and 
regulations that apply to the LNG industry further ensure safety. 

\begin{figure}[tb]
\centering
\epsfig{file=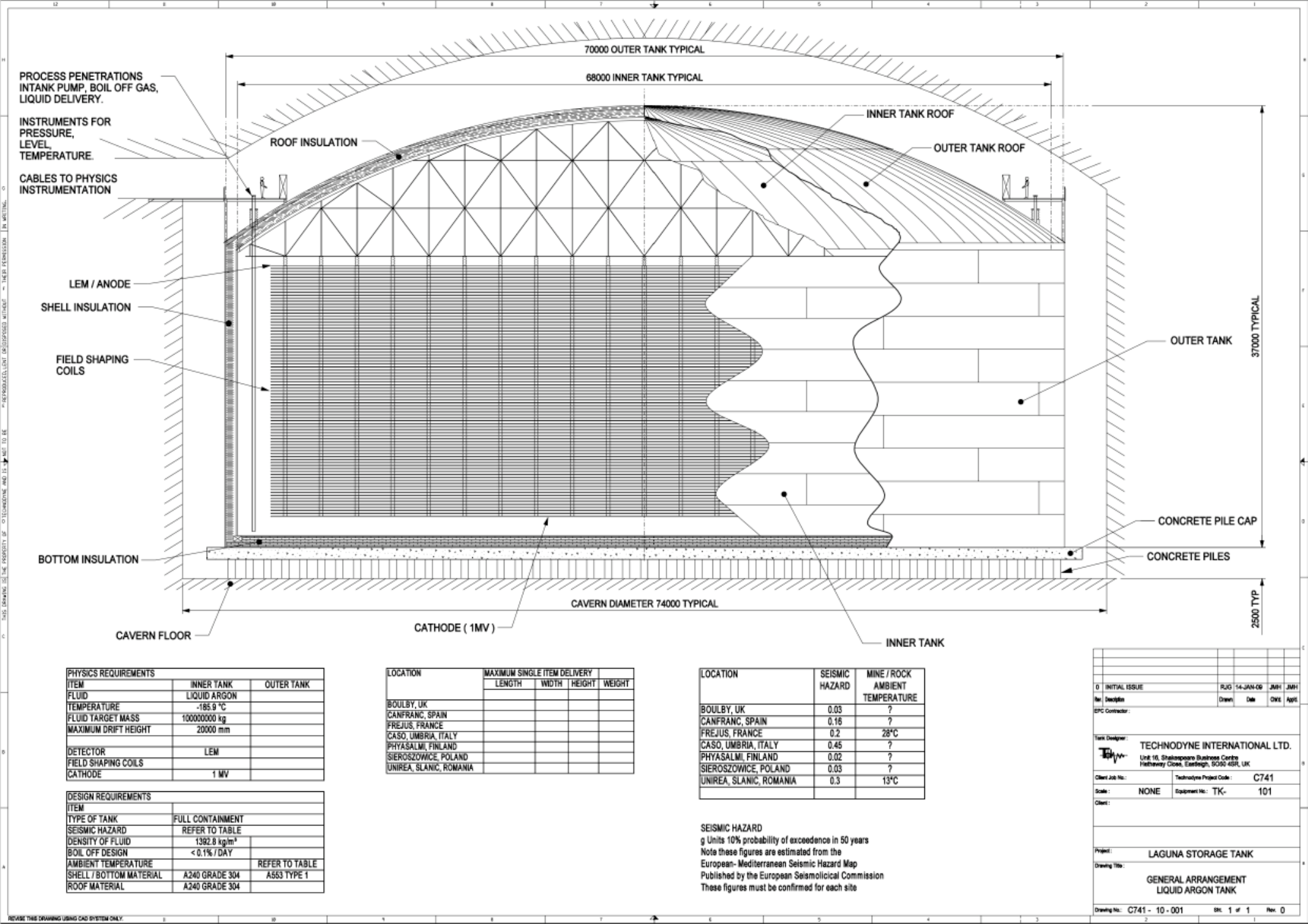,width=\textwidth}
\caption{Design of a 100~kton liquid Argon GLACIER tank developed
in Collaboration with Technodyne International Limited~\protect\cite{Technodyne}.}
\label{fig:glacierdesign}
\end{figure}

The unit volume cost of full containment LNG surface tanks was about $ \$400/m^3$ in 2003 and culminated
to about twice that in 2008. Larger tanks are generally cheaper per unit volume. Single containment
tanks are about half the price. Membrane tanks have been historically less well accepted, hindered
by lack of accepted standards. In 2006  the European Standard EN~14620 developed to
include LNG membrane technology, which could therefore
become a serious alternative to single or full containment large tanks.

Already in 2004 we have appointed the Technodyne International Limited, UK~\cite{Technodyne}, an expert
company in the design of LNG tanks, to initiate a feasibility
study in order to understand and clarify the issues related to the operation of a large
underground LAr detector.  
The preliminary Technodyne study was sufficient to understand that LNG storage tank could
be adequately extrapolated to the case of liquid Argon.  The boiling points of LAr and CH4 
are 87.3 and 111.6~K and their latent heat of vaporization per unit volume is the same for both liquids 
within 5\%. The main differences are that  (a)
 LNG is flammable when present in air within 5 to 15\% by volume while LAr is not flammable
(b) the density of  LAr is 3.3 times larger than LNG, hence the tank needs to withstand 
higher hydrostatic pressure. This is taken into account in the engineering design.
The estimated boil-off in the tank is 0.04\%/day. Other cryogenic aspects have been 
discussed in Ref.~\cite{Rubbia:2004tz}.

 For the requirements the recommended 
design -- See Figure~\ref{fig:glacierdesign} -- is the full containment composed of an inner and an outer tank made from steel
and of the following  principal components: 
\begin{enumerate}
\item A 1m thick reinforced concrete base platform 
\item Approximately one thousand 600mm diameter 1m high support pillars 
arranged on a 2m grid. Also included in the support pillar would be a seismic / 
thermal break. 
\item A 1m thick reinforced concrete tank support sub-base. 
\item An outer tank made from stainless steel, diameter 72.4m. The base of which 
would be approximately 6mm thick.  The sides would range from 48mm thick 
at the bottom to 8mm thick at the top. 
\item 1500mm of base insulation made from layers of felt and foamglas blocks. 
\item A reinforced concrete ring beam to spread the load of the inner tank walls. 
\item An inner tank made from stainless steel, diameter 70m. The base of which 
would be approximately 6mm thick and the sides would range from 48mm thick 
at the bottom to 8mm thick at the top. 
\item A domed roof with a construction radius of 72.4m attached to the outer tank 
\item A suspended deck over the inner tank to support the top-level instrumentation 
and insulation.  This suspended deck will be slightly stronger than the standard 
designs to accommodate the physics instrumentation.  This in turn will apply 
greater loads to the roof, which may have to be strengthened, however this is 
mitigated to some extent by the absence of wind loading that would be 
experienced in the above ground case. 
\item Side insulation consisting of a resilient layer and perlite fill, total thickness 
1.2m. 
\item Top insulation consisting of layers of fibreglass to a thickness of approximately 
1.2m. 
\end{enumerate}

Within LAGUNA, a more detailed design is being developed 
in Collaboration with Technodyne. In addition to the tank,
the interface between tank and detector is
being addressed.

\begin{figure}[htb]
\centering
\epsfig{file= 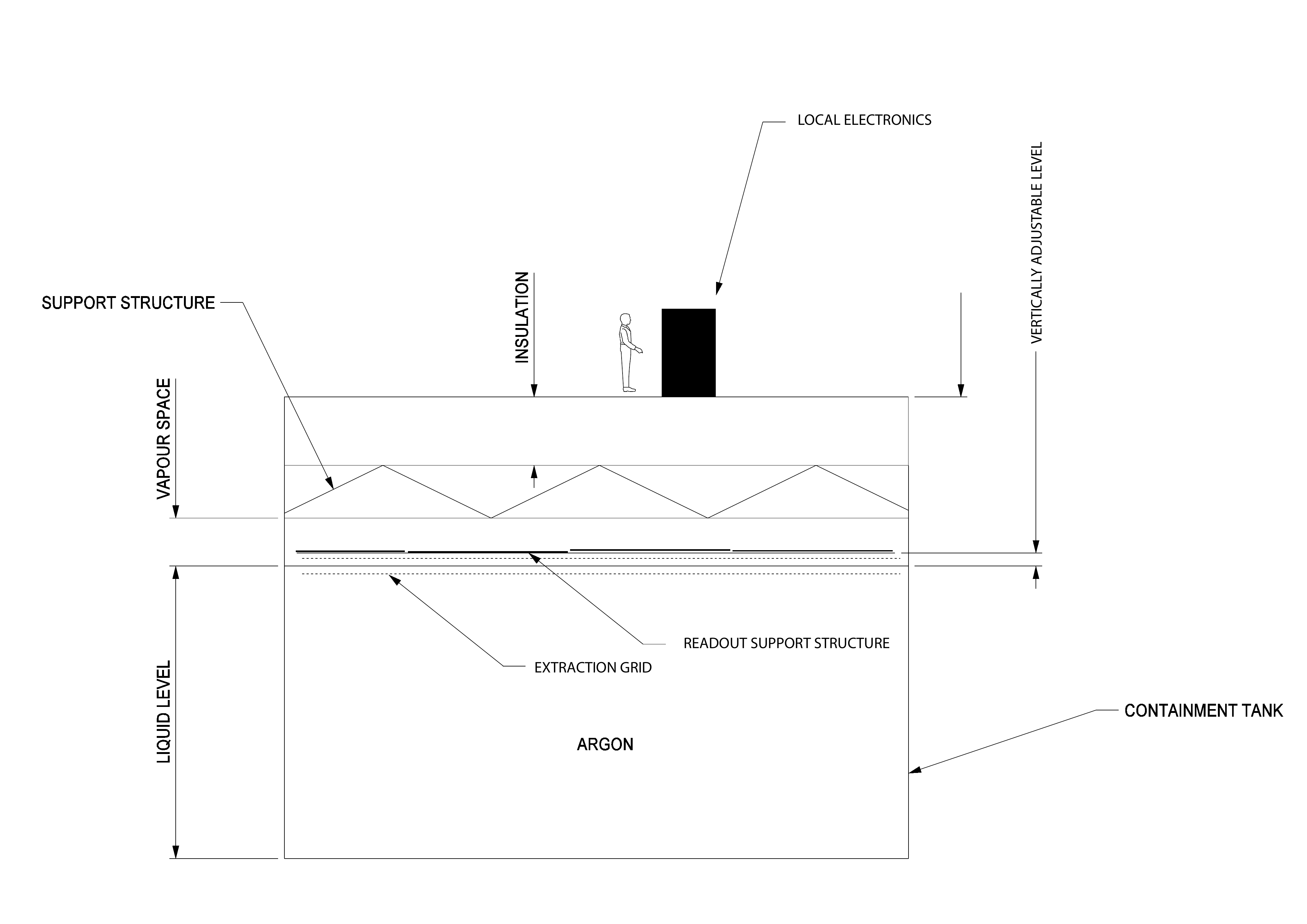,width=\textwidth}
\caption{GLACIER concept for readout based on double phase operation (not to scale).}
\label{fig:glacierreadout}
\end{figure}

\subsection{A very large area LAr LEM-TPC}

GLACIER is based on a novel concept 
for readout called LAr LEM-TPC~\cite{Badertscher:2008rf}. 
LAr LEM-TPCs operate in double phase with charge extraction and amplification in the vapor phase -- see Figure~\ref{fig:glacierreadout}. 
The concept has been very successfully demonstrated on small prototypes: the electron drift direction
is chosen vertical for full active volume; ionization electrons, after drifting in the LAr volume, are 
extracted by a set of grids into the gas phase and driven into the holes of a double stage Large Electron Multiplier (LEM), where charge amplification occurs. Each LEM is a thick macroscopic hole multiplier, which can be manufactured with standard PCB techniques. The electrons signal is readout via two orthogonal coordinates, one using the induced signal on the segmented upper electrode of the LEM itself and the other by collecting the electrons on a segmented anode. The images obtained with the LAr LEM-TPC are of very high -- ``bubble-chamber-like'' -- quality, owing to the charge amplification in the LEM and have good measured dE/dx resolution. Compared to LAr TPCs with immersed wires, whose scaling is at least limited by mechanical and capacitance issues of the long thin wires and by signal attenuation along the drift direction, the LAr LEM-TPC is an elegant solution for very large liquid Argon TPCs with long drift paths and mm-sized readout pitch segmentation. Low detection thresholds are possible and the method could in principle be generalized to
a 3D pixelized readout.

\begin{figure}[tb]
\begin{minipage}{20pc}
\includegraphics[width=20pc]{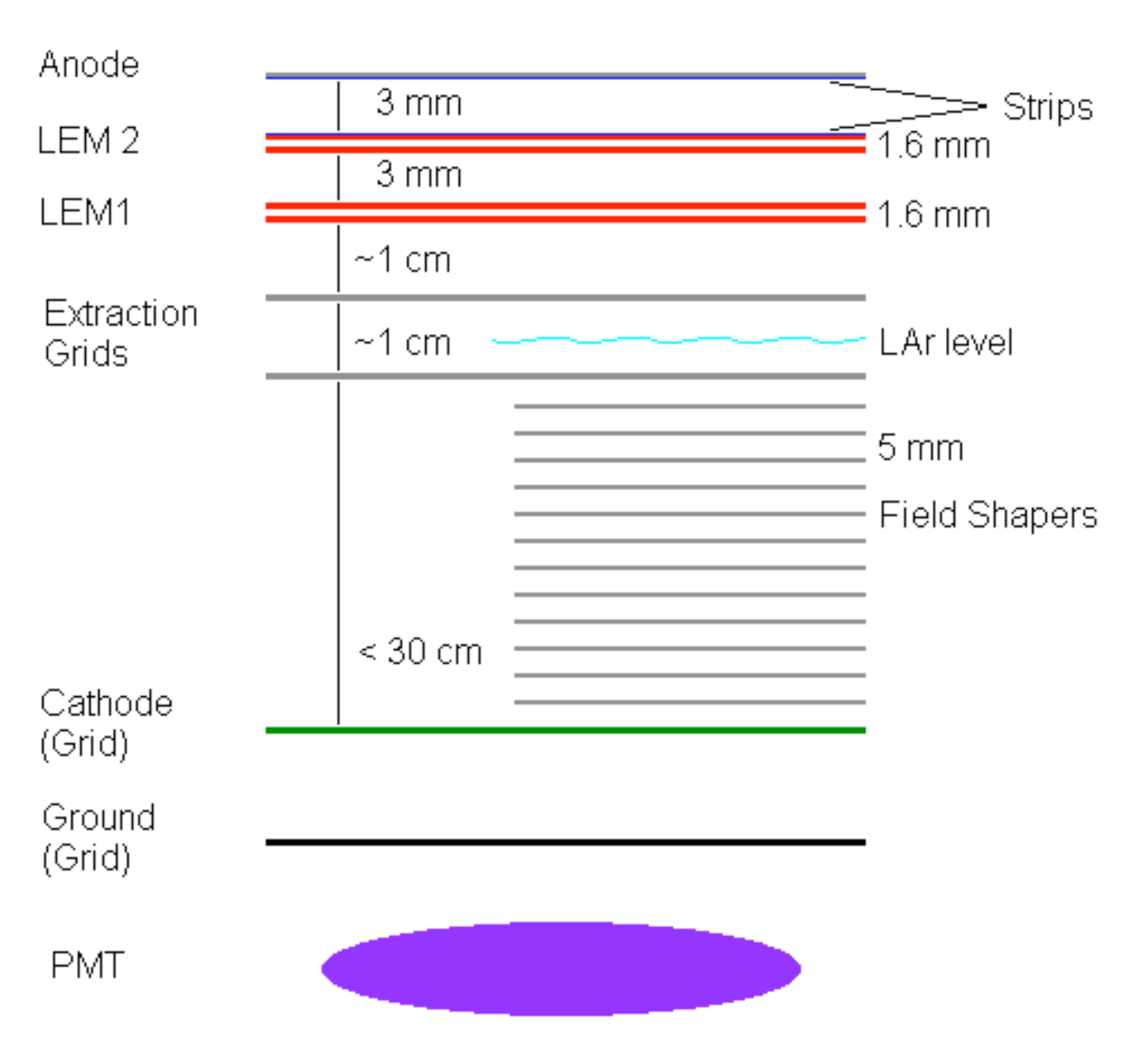}
\caption{\label{LEMScheme}Schematic of a LEM-TPC setup showing the LAr level between the two extraction grids. When operated
 in pure Ar gas, radioactive sources were placed on the ground grid above the PMT. From Ref.~\cite{Badertscher:2008rf}.}
\end{minipage}\hspace{2pc}%
\begin{minipage}{11pc}
\includegraphics[width=11pc]{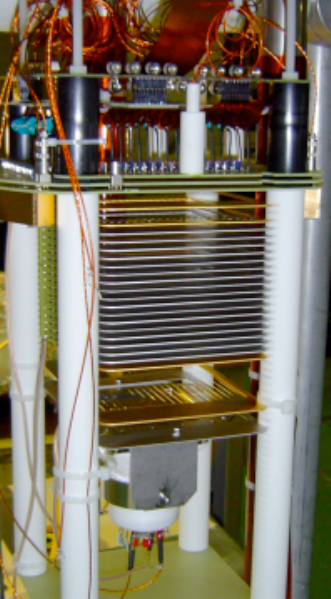}
\caption{\label{LEM-TPCSetup}Assembly of a LAr~LEM-TPC prototype. From Ref.~\cite{Badertscher:2008rf}.}
\end{minipage} 
\end{figure}

The general technique of electron multiplication via avalanches in small holes
is attractive because (1)~the required high electric field can be naturally attained
inside the holes and (2)~the finite size of the holes effectively ensures 
a confinement of the electron avalanche, thereby reducing secondary 
effects in a medium without quencher. 
The gain ($G$) in a given uniform electric field of a parallel plate chamber  
at a given pressure is described by $G\equiv~e^{\alpha d}$ where 
$d$ is the gap thickness and  $\alpha$
 is the Townsend coefficient, which represents the number of electrons
 created per unit path length by an electron in the amplification region. 
The behavior of this coefficient 
with pressure and electric field can be approximated by the Rose and 
Korff law~\cite{RoseKorff}: $\alpha=A\rho~e^{-B\rho~/E}$
where  $E$ is the electric field, 
$\rho$ is the gas density, $A$ and $B$ are the parameters depending on the gas.
Electron multiplication
in holes has been investigated for a large number of applications.
The most extensively studied device is the  Gas Electron Multiplier~(GEM)~\cite{Sauli97}, 
made of
50--70~$\mu$m diameter holes etched in a 50~$\mu$m thick metalized
Kapton foil.  Stable operation has been shown with various gas mixtures
and very high gains.
An important step was the operation of the GEM in 
pure Ar at normal pressure and temperature~\cite{Sauli99}. 
Rather high gas gains were obtained, of the order of 1000, supporting  evidence
for the avalanche  confinement to the GEM micro-holes.
The successes of the GEMs triggered the concept of the LEM or 
THGEM (for a recent review see~\cite{Bondar08}), a coarser but more rigid
structure made with holes of the millimeter-size in a millimeter-thick
printed circuit board~(PCB).

In order to study the properties of the LEM and the possibility to
reach high gains in double phase, we have performed extensive R\&D on several prototypes~\cite{Polina_ETHthesis}:
We have built several LEM prototypes using standard PCB techniques from different manufacturers.
Double-sided copper-clad (16~$\mu$m layer) FR4 plates with thicknesses ranging from 0.8~mm to 1.6~mm
are drilled with a regular pattern of 500~$\mu$m diameter holes  at a relative distance of 800~$\mu$m. 
By applying a potential difference on the two faces of the PCB an intense electric field inside the holes is produced.

\begin{figure}[tb]
\includegraphics[width=14pc]{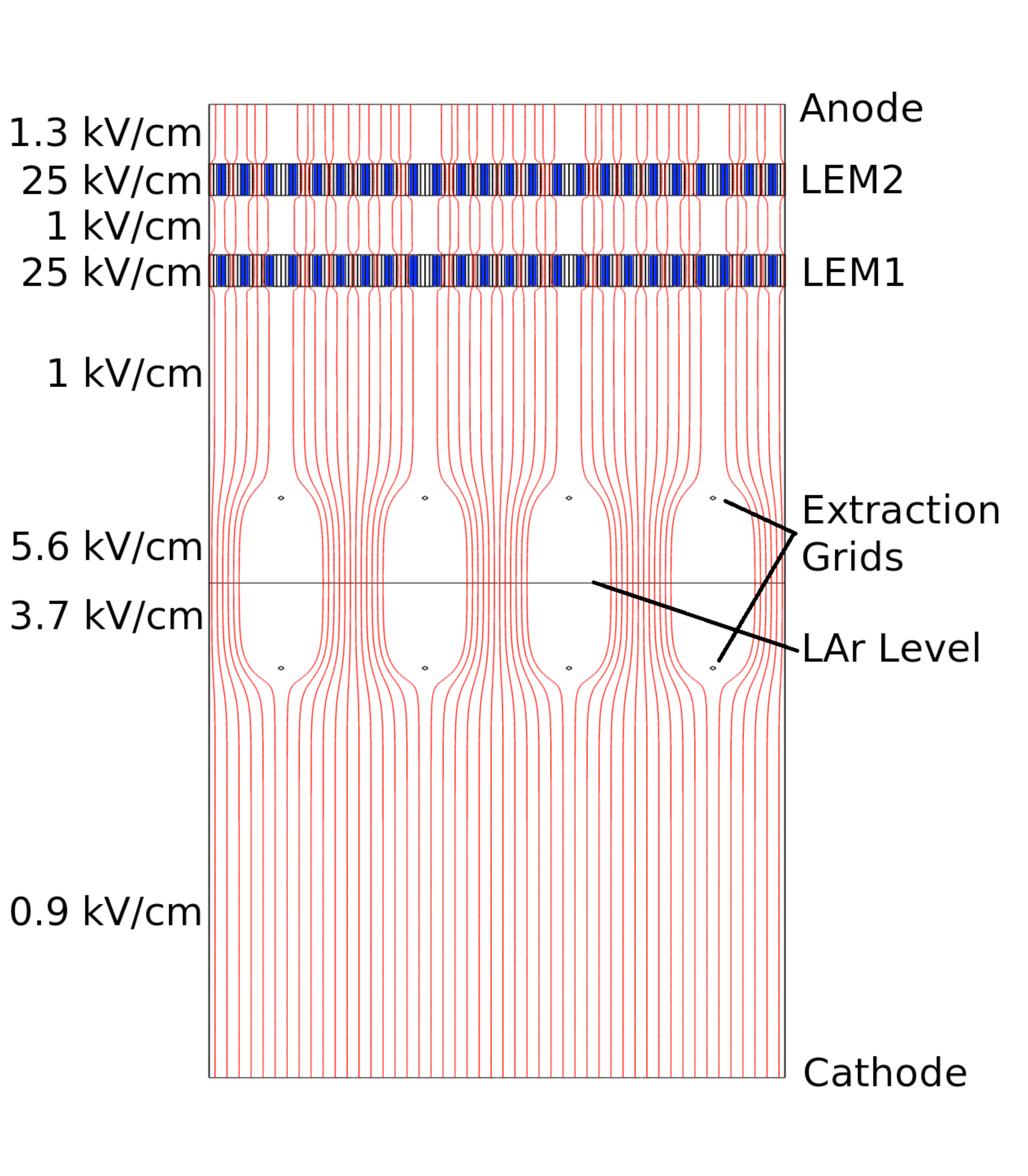}\hspace{2pc}%
\begin{minipage}[b]{20pc}\caption{\label{fieldLines}Electric field lines in the double phase operation.  From Ref.~\cite{Badertscher:2008rf}.}
\end{minipage}
\end{figure}

A first single stage prototype demonstrated a stable operation in pure Ar 
at room temperature and pressure up to 3.5~bar with a gain of 800 per electron. 
Measurements were performed at high pressure because the density of Ar at 3.5~bar is roughly
equivalent to the expected density of the vapour at the temperature of 87~K. Simulations 
of the LEM operation were performed using the MAXWELL (field calculations) and 
MAGBOLTZ (particle tracking) programs. The results obtained were in good agreement 
with the experiment. This suggested
the use of the same formalism of the parallel plate chamber by replacing the gap thickness $d$
by the effective amplification path length within the holes, called $x$, which can be estimated
with electrostatic field calculations as the length of the field plateau along the hole.
The gain is then expressed as  $G_{LEM} = e^{\alpha x}$, where $\alpha$
is the first Townsend coefficient
at the maximum electric field $E$ inside the holes.
For example,  simulations indicate that $x\simeq 1$~mm
for a LEM thickness of 1.6~mm, hence, $\alpha(cm^{-1})\simeq\ln G/(0.1~cm)$.

Double stage LEM configurations were tested in pure Ar at room
temperature, cryogenic temperature and in double phase conditions.
Tests in an Ar/CO$_2$ (90\%/10\%)
gas mixture were also performed to compare the results with those obtained in pure Ar.
The double-stage LEM system demonstrated a gain of $\sim$10$^3$ at a temperature of 87~K
and a pressure of $\sim$1~bar. The double-phase operation of the LEM proved the extraction
of the charge from the liquid to the gas phase.
\begin{figure}[hbt]
\centering
\includegraphics[width=0.95\textwidth]{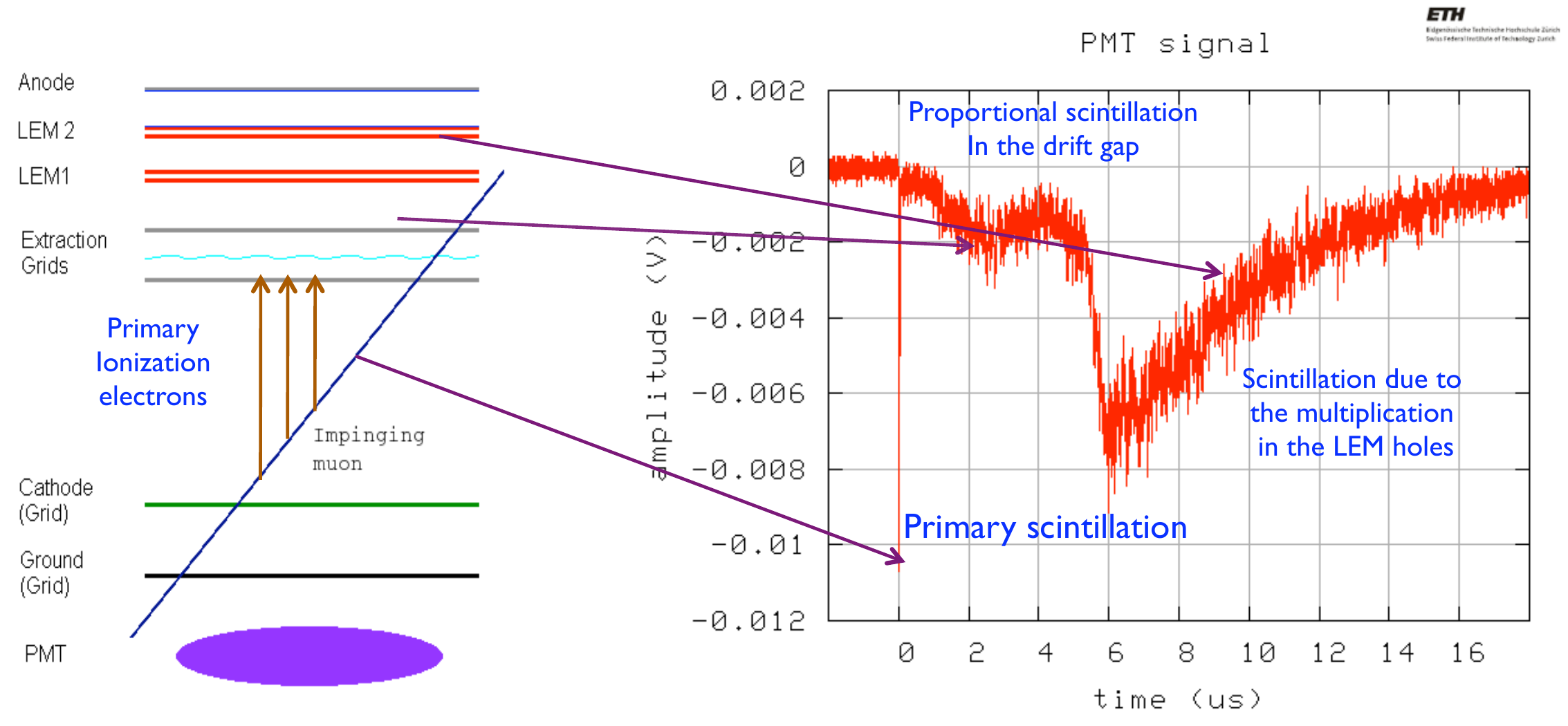}
\caption{Typical signal waveform of the immersed PMT Hamamatsu R6237-01.}
\label{fig:pmtsignal}
\end{figure}

A $\sim$3~lt active volume LAr LEM-TPC, as shown 
schematically in Figure~\ref{LEMScheme}, was constructed and successfully operated.
A LAr drift volume of 10x10~cm$^2$ cross section and with an adjustable 
depth of up to 30~cm is followed on top by a double stage LEM positioned in the Ar vapour at about 1.5~cm 
from the liquid. Ionization electrons are drifted upward by a uniform electric field generated by a system of 
field shapers, extracted from the liquid by means of two extraction grids
positioned across the liquid-vapour interface and driven onto the LEM planes.
The extraction grids were constructed as an array of parallel stainless steel wires of 100~$\mu$m diameter with 5~mm spacing.
A cryogenic photomultiplier (Hamamatsu R6237-01) is positioned below the drift region and electrically 
decoupled from the cathode at high voltage by a grid close to the ground potential. The photomultiplier is coated with
tetraphenylbutadiene~(TPB) that acts as wavelength shifter for 128~nm
photons of Ar scintillation. 

\begin{figure}[tb]
\centering
\includegraphics[width=\textwidth]{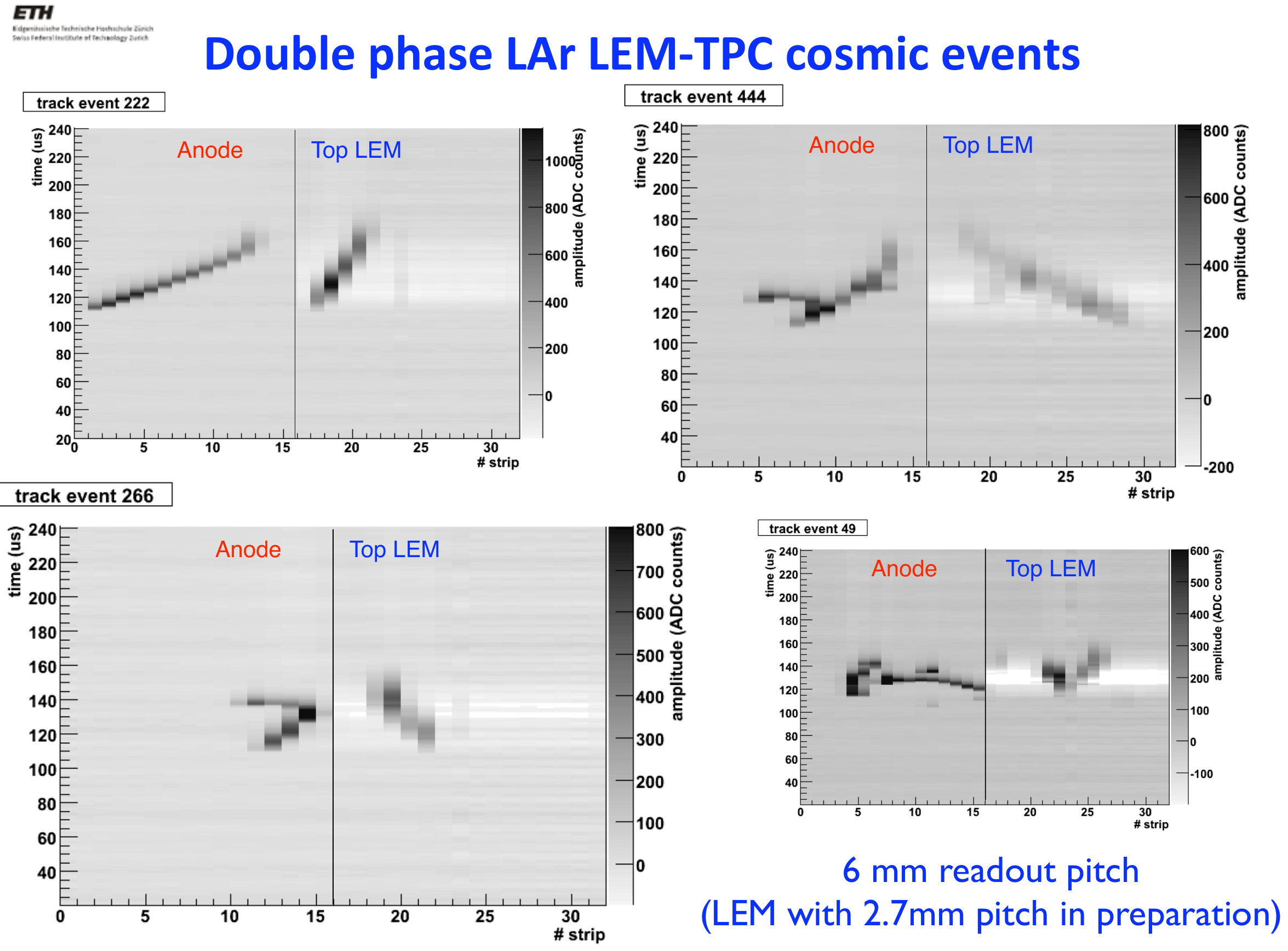}
\caption{Display of a typical cosmic ray event in double phase operation. Channels 0-15 are connected to LEM strips and channels 16-31 to anode strips.}
\label{liquidEvent}
\end{figure}

Ionization electrons undergo multiplication into a first LEM plane and the
resulting charge is then driven into a second LEM plane for further
multiplication.
The amplified charge is readout by measuring two orthogonal coordinates, one using the induced signal 
on the segmented upper electrode of the second LEM itself and the other by collecting the electrons on a segmented anode. 
In this first production both readout planes are segmented with 6~mm wide strips, for a total of 32 readout channels for a $\sim$10x10~cm$^2$ active area.
Transverse segmentations down to 2--3~mm will be tested in the near future.

Signals from LEM and anode strips are decoupled via high voltage capacitors and routed to a signal
collection plane placed a few centimeter above the anode. Each signal line is equipped with a surge
arrester to prevent damaging the preamplifiers in case of discharges.
The detector is housed inside a vacuum tight dewar and Kapton flex-print are used to connect the
signal lines on the signal collection board to the external readout electronics.
The flex-prints exit the dewar through a slot cut in an UHV flange and sealed with a cryogenic
epoxy-resin to maintain vacuum tightness.

Figure~\ref{fieldLines} shows the electric field lines from the
cathode to the anode for the double phase operation. The electric fields are set increasingly from
the drift region towards the anode such that fields lines starting
at the cathode reach the anode (transparency).

In double phase operation the device gain was set to about 10.
In this mode of operation, the PMT signal proved to be very useful in order to analyse the faith of primary ionization electrons (See Figure~\ref{fig:pmtsignal}): 
a fast light peak indicated the direct scintillation of the crossing cosmic muon. A second light peak, shifted in time if the ionizing
track did not cross the liquid surface, corresponded to the proportional scintillation (luminescence) in the high field region in the
vapour just above the liquid. A third peak was interpreted as light produced during the multiplication avalanche, which escapes the
LEM holes. Indeed, the size of the 3rd light peak was correlated with the electric field inside the LEM.
An example of cosmic muon track is shown in Figure~\ref{liquidEvent}. This represents a proof of principle of the operation of a double phase LAr~LEM-TPC as a tracking device.

\subsection  {A light readout system for triggering and also Cerenkov light imaging}
Ionizing radiation in liquid noble gases leads to 
the formation of excimers in either singlet or triplet states~\cite{Doke:1990p1608,Kubota:1978p975}, which decay radiatively
to the dissociative ground state with characteristic fast and slow lifetimes  ($\tau_{fast}\approx 6$~ns, $\tau_{slow}\approx 1.6\mu$s in liquid argon
 with the so-called second continua emission spectrum peaked at 
$128\pm10$~nm~\cite{larlight1}).
Scintillation light readout provides information on events occurring
in the detector. The correlation between detected light and the arrival of the charge furnishes the $T_0$
of the event, defining the coordinate along the drift direction. For pulsed beam
events, the triggering allows to suppress backgrounds. 

Large VUV sensitive PMTs (e.g.\,MgF$_2$ windowed) 
are not commercially available, the use of reflectors coated with a wavelength shifter (WLS) 
along with standard bialkali photomultiplier tubes (PMTs) is an
economical realisation of an efficient readout system~\cite{Boccone:2009kk}. With about 40'000~$\gamma$'s emitted
per MeV deposited (at zero field) the number of PMTs does not need to be very large. We consider
that about 1000~8''~PMTs will be sufficient for the 100~kton GLACIER.

Liquid Argon has very similar Cerenkov light emission
properties than water and also similar physical properties in terms of radiation lengths,
interaction lengths, etc.. hence the events in liquid Argon look very
much the same as those in water and the techniques developed 
for the reconstruction and analysis of Water Cerenkov detectors could in principle be ``transposed''
to the liquid argon case. 

\begin{figure}[h]
\centering
\epsfig{file=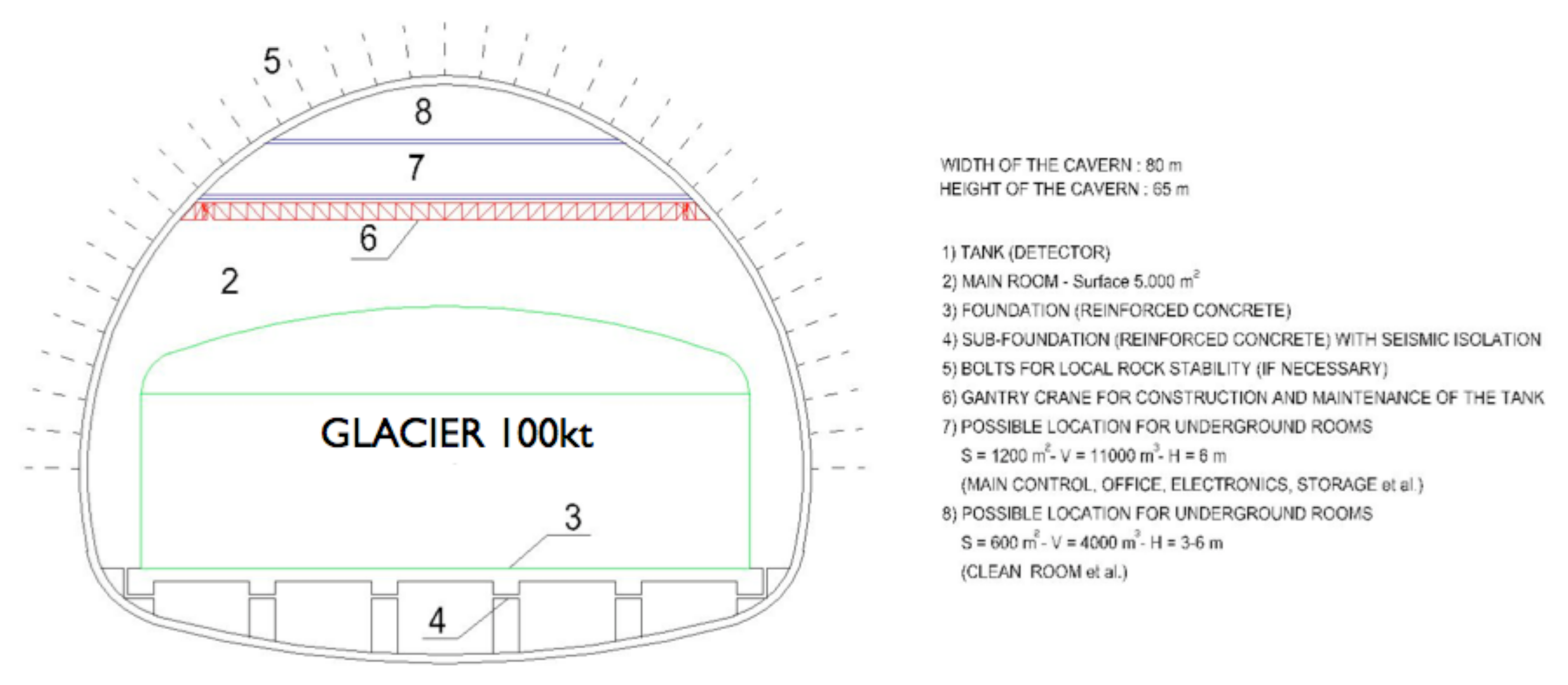,width=\textwidth}
\caption{Cavern layout developped by AGT Ingegneria~\protect\cite{AGT}.}
\label{fig:agtshallowpdepth}
\end{figure}

\subsection {The underground localization} 
A detector like GLACIER should be located underground in order
to provide the best possible physics output given its mass. With a shallow
depth, cosmic muons induced reactions 
which will provide dangerous backgrounds
to rare events. We are currently studying within the LAGUNA design study two possible configurations: (1) a cavern
in a mine (with a vertical access through a shaft) (2) a cavern in a mountain (with
a horizontal access through a tunnel).
As an example, a preliminary project developed in Collaboration with AGT
Ingegneria~\protect\cite{AGT} is shown in Figure~\ref{fig:agtshallowpdepth}.

\subsection{The possibility of a magnetic field:} 
The possibility to complement the features of GLACIER
with those provided by a magnetic field has been considered in
the past~\cite{Bueno:2001jd,Rubbia:2001pk} and would open new
possibilities: (a) charge discrimination, (b)
momentum measurement  ($e.g.$ high energy muons), (c)
precise kinematics, since the measurement precision is limited by multiple
scattering ($e.g.$ $\Delta p/p\approx$4\% for a track length of $L\!=\!12$\,m and a field  of
$B\!=\!1$\,T). 
The orientation of the magnetic field can be chosen such that the bending direction is
in the direction of the drift where the best spatial resolution is achieved. This is
possible since the Lorentz angle is small~\cite{Badertscher:2004py,Badertscher:2005te}. However, this is not
mandatory and the B--field could also be parallel to the drift field.

\begin{table}[htb]
\caption{Comparison of superconducting solenoidal magnets. ATLAS column corresponds
to the solenoid. From Ref.~\cite{Ereditato:2005yx}.}
\label{table:magpars}
\begin{tabular}{@{}|l|c|c|c|c|}
\br
	& 10 kton LAr	&		100 kton LAr		&	ATLAS & CMS \\
	\mr
Magnetic induction (T) & 0.1/0.4/1.0  & 0.1/0.4/1.0 &2.0&4.0 \\
Solenoid diameter (m) &30	&		70		&	2.4	&6\\
Solenoid length (m) &10		&	20	&		5.3&	12.5\\
Magnetic volume (m$^3$)&	7700	&		77000	& 21& 400\\
Stored magnetic energy (GJ)&	0.03/0.5/3 &	0.3/5/30	&0.04 & 2.7\\
Magnetomotive force (MAt)&	0.8/3.2/8 & 1.6/6.4/16	&9.3 &42\\
Radial magnetic pressure (kPa)&	4/64/400	& 4/64/400 & 1600&  6500\\
Coil current (kA) & 	\multicolumn{2}{c|}{30   (I/Ic=50\%)}		& 8 & 20\\
Total length conductor (km)&	2.5/10/25	& 12/57/117& 5.6 & 45\\
Conductor type	& \multicolumn{2}{c|}{NbTi/Cu superconductor} & NbTi/Cu& NbTi/Cu\\
& \multicolumn{2}{c|}{or HTS superconductor ? (see text)} & & \\
\br
\end{tabular}
\end{table}

The presence of magnetic field is certainly
beneficial for the application in the context of the neutrino factory,
depending on the actual field strength~\cite{Rubbia:2001pk}: 
(1) a low field, $e.g.$ B=0.1~T, for the measurement of 
the muon charge (CP-violation); 
(2) a strong field, $e.g.$ B=1~T for the measurement of the muon/electron charges (T-violation).
For a practical application, however, the mass of the detector will have to be very large,
in the multikton range.

We have successfully operated the first experimental prototype of a magnetized 
liquid Argon TPC~\cite{Badertscher:2004py,Badertscher:2005te}.
Beyond the basic proof of principle, the main challenge to be addressed is
the possibility to magnetize a very large mass of Argon, in a range of 10~kton or more.
The most practical design, which fits the GLACIER concept,
is that of a vertically standing solenoidal coil producing vertical field lines, parallel
to the drift direction.
Hence, to magnetize the very large LAr volume one could immerse a superconducting 
solenoid directly into the LAr tank to create a magnetic field, parallel to the drift-field~\cite{Ereditato:2005yx}.

\begin{table}[tb]
\caption{Tentative parameters for return iron yoke. From Ref.~\cite{Ereditato:2005yx}.}
\label{table:magyoke}
\centering
\begin{tabular}{@{}lccc}
\br
Cyl. Fe yoke &	10 kton LAr		&	100 kton LAr\\
\mr
Magn. ind. (T)&	0.1/0.4/1.0	&0.1/0.4/1.0\\
Magn. flux (W)&	70/280/710	&385/1540/3850\\
Field in Fe (T)&	1.8		&	1.8		\\
Thickness (m)&	0.4/1.6/3.7	&1/3.7/8.7\\
Height (m)&	10		&	20		\\
Iron Mass (kton)	&6.3/25/63 &34/137/342\\
\br
\end{tabular}
\end{table}

In Table~\ref{table:magpars} we summarize the relevant physical parameters 
for a 10 and 100~kton liquid Argon detectors for three different magnetic
field configurations (0.1, 0.4 and 1.0~T), compared
to existing solutions for LHC experiments.  Of course, a very large amount of energy will be stored
in the magnetic fields. However, owing B$^2$ dependence of the energy density,
the total amount of magnetic energy is comparable to that of
LHC experiments. For example, the magnetic volume of
a 10~kton liquid Argon detector is 7700~m$^3$, 
to be compared to 400~m$^3$ for CMS,
 $i.e.$ an increase by a factor $\simeq 20$. However, the CMS
field is 4~T, hence,  a 10~kton LAr detector with 
a field B$\approx 4/\sqrt{20}\approx 1~$T has
the same stored energy.
Since most issues related to large magnets 
(magnetic, mechanical, thermal) scale with the stored energy, we
readily conclude, that  the magnetization
of very large liquid Argon volumes, although certainly very challenging,
is not unrealistic! More figures of comparison
are shown in Table~\ref{table:magpars}.

In the case of quenching, 
the liquid Argon also plays the role of the thermal bath.
In the largest considered volume, $i.e.$ for a 100~kton detector, and
for a B=0.1T (resp. 1T), the stored energy in the B-field is 280 MJ (resp. 30 GJ). 
In case of full quenching of the coil, the LAr would absorb the dissipated heat 
which would produce a boil-off of 2 tons (resp. 200 tons) of LAr. 
This corresponds to 0.001\% (resp. 0.1\%) of the total LAr contained in the tank 
and hence supports the design.
			
A magnetized LAr TPC was already addressed by Cline et al.~\cite{Cline:2001pt}.
However, the presence of long wires disfavored the use of a magnetic field. 
In addition, the proposed warm coil would dissipate 17~MW at B=0.2~T (88 MW @ 1~T).
Such heat, even if affordable, would impose strong technical constraints
on the thermal insulation of the main tank. 
In contrast, superconductors produce no heat dissipation and the coil current 
flows even in absence of the power supply.

\subsubsection {The use of High $T_C$ superconductors?}
A new era in superconductivity opened in 1986 when Bednorz and Mueller in Zurich discovered superconductivity at a temperature of approximately 30~K. In the following years, new materials were found and currently the world record is T$_c\approx$130~K.
HTS are fragile materials and are still at the forefront of material science research. 
However, they hold large potentials for future large scale applications.
For example, ribbons of BSCCO-2212 ($Bi_2-Sr_2-Ca_2-Cu_3-O_{10}$) with $T_c=$110 K are 
very promising and are regularly produced by $e.g.$ American 
Superconductor~\cite{amsuper}. 
Tapes of Bi2223 or YBCO are promising 2nd generation
HTS cables which will find many
applications in the next years. Unfortunately, HTS superconductors are much
more difficult to use than normal superconductor. In particular, their conductivity
depends strongly on temperature and on external perpendicular magnetic fields.
Nonetheless, magnets have been constructed with HTS.
The development of  Superconducting Magnetic Energy Storage (SMES)
systems will surely drive the technical developments of these HTS ribbons
in the years to come.
Today it is conceivable to estimate that very large coils from 
BSCCO-2212 could be operated at T=10~K,
with BSCCO-2223 @ T=20~K, and possibly YBCO @ T=50~K~\cite{strauss}.

\subsubsection {Tentative yoke parameters}

In order to close the magnetic field lines, one can think of a simple
cylindrical and hollow iron ring to encompass the liquid Argon
volume. The thickness of the iron ring is determined by the magnetic
flux that needs to be absorbed. We summarize in Table~\ref{table:magyoke} the 
relevant dimensions.

In the case of SMES considered for underground storage of MJ energy,
systems without return yoke buried in tunnels in bedrock have
been considered and studied~\cite{eyssa}. If an underground location where stray
field can be tolerated is considered, then one could avoid using a yoke.

%
%

\begin{figure}[tb]
\centering
\epsfig{file=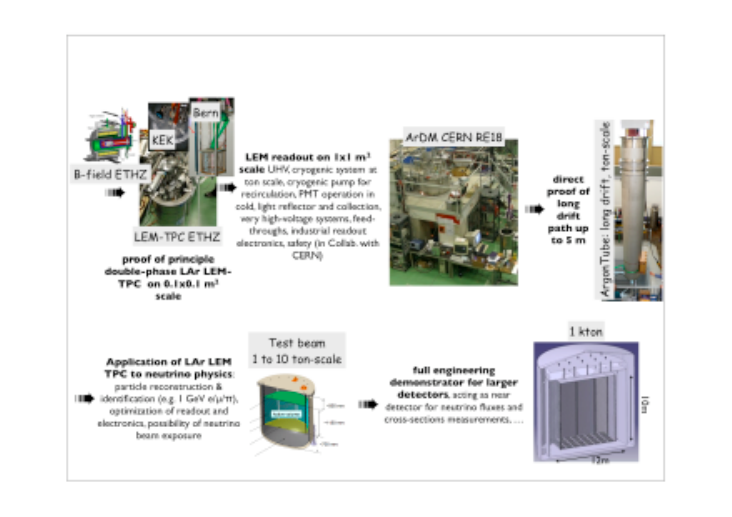,width=\textwidth}
\caption{The ``graded'' strategy towards GLACIER.}
\label{fig:glacierstrat}
\end{figure}

\section{Further R\&D steps towards GLACIER}
Since 2004, several critical issues have been identified and are subject to 
intense R\&D efforts towards GLACIER, which are outlined below:
\begin{enumerate}
\item the tank is being further studied and designed with the
engineering company Technodyne within the LAGUNA design
study;
\item the readout system based on the novel technique of the
LAr LEM-TPC is being further studied and perfected on larger
area. A $40\times 40$~cm$^2$ LEM is being
assembled and tested;
\item the optimization for very long drift paths will be the subject
of several investigations and  tested in dedicated test (ArgonTUBE);
\item the drift HV system based on a Cockroft-Walton voltage multiplier
has been successfully tested on a prototype immersed in LAr and a 500~kV
circuit will be further tested 
in the ArDM experiment;
\item new modern solutions for electronic readout
have been developed in collaboration with the CAEN company.
In addition, an ASIC version of a preamplifier has been designed and a prototype manufactured.
Cold and warm solutions are being worked on~\cite{ipnl}. An ethernet based
readout chain including a network time distribution is undergoing
tests;
\item large scale liquid purification is being developed in 
collaboration with industry;
\item the safety of GLACIER is being addressed in a work package
of the LAGUNA design study;
\end{enumerate}

The graded strategy  is 
summarized in Figure~\ref{fig:glacierstrat}. It is composed
of several steps:
\begin{enumerate}
\item small laboratory setups (already achieved);
\item the ArDM-1t experiment (ongoing);
\item the ArgonTUBE dedicated to the study of 5~m drift (preparation);
\item a test beam campaign specific to the application
to neutrino physics (under investigation);
\item a 1~kton full engineering demonstrator acting as near detector for neutrino fluxes and cross-section measurements (under investigation).
\end{enumerate}

\subsection{The CERN RE18 (ArDM) experiment}

\begin{figure}[tb]
\centering
a)\includegraphics[width=0.4\textwidth]{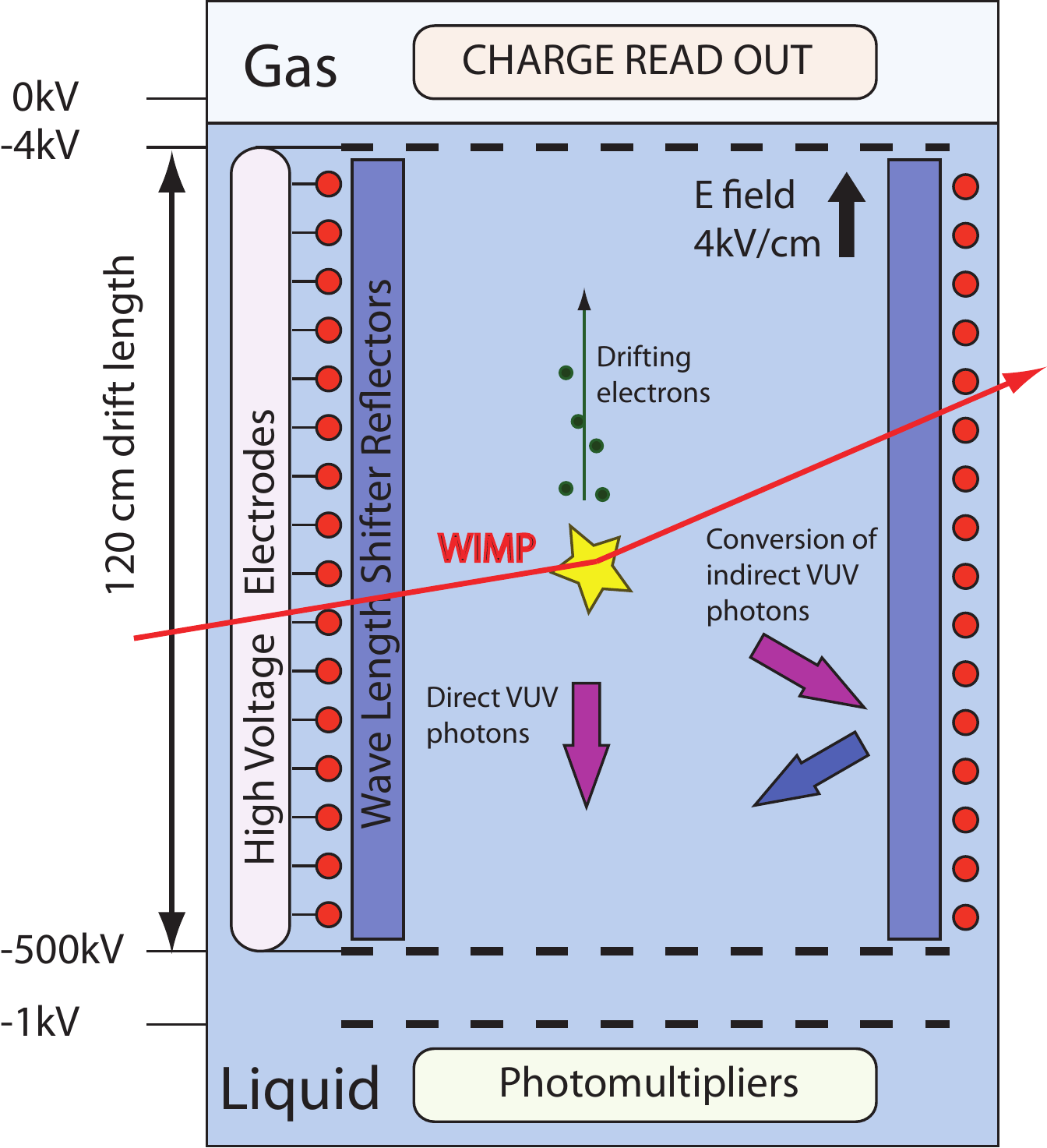}\hspace{18mm}b)\includegraphics[width=0.4\textwidth]{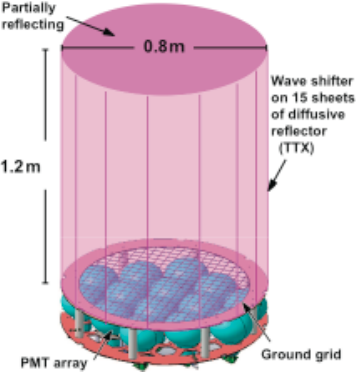}
\caption{(left) Operation principle of the ArDM detector, (right) 3D sketch of the  83\,cm by 120\,cm cylindrical light detection volume.
From Ref.~\cite{Boccone:2009kk}.}
\label{fig:intro}
\end{figure}

The ArDM experiment~\cite{Rubbia:2005ge},
presently under construction at CERN (RE18),
aims for the operation of a ton-scale liquid argon 
target for direct detection of dark matter.
The detector is operated as a double phase LAr LEM-TPC~\cite{Badertscher:2008rf}, as is illustrated in Figure\,\ref{fig:intro}(left). 
The charge readout is situated in the gas phase at the top of the vessel, while photodetectors sensitive to single photons are located 
in the liquid argon at the bottom of the apparatus.
The light readout system is  composed of 14 hemispheric  8'' Hamamatsu R5912-02MOD-LRI
PMTs, made from particularly radiopure 
borosilicate glass and feature bialkali photocathodes with Pt-underlay 
for operation at cryogenic temperatures~\cite{Bueno:2007au}. 
The argon scintillation light  is 
wavelength shifted into the range of maximum quantum efficiency (QE) of the PMT
by a thin layer~(1.0~$\mgcs$) of tetra\-phenyl\-buta\-diene 
(TPB) evaporated onto 15 cylindrically arranged reflector sheets 
(120$\times$25\,cm$^2$) which are located in the vertical electric field. 
These sheets are made of the PTFE fabric Tetratex{\small\textregistered} 
(TTX), transporting the shifted light diffusively onto the PMTs 
(shown in Figure\,\ref{fig:intro} (right)). 
The PMT glass windows are also coated with TPB to convert directly 
impinging DUV photons. 

The ArDM-1t detector is being assembled at CERN. After
successfully completing the tests of the detector and the calibration of
its response to low energy particles, an underground location will be considered.

\subsection{ArgonTUBE}
In order to experimentally assess very long drift path, 
a full  scale measurement of long drift paths (5 m) is currently being designed and assembled~\cite{Ereditato:2005ru}.
This tests aimed at dedicated studies of signal attenuation and multiplication
and of effect of charge diffusion will simulate Ôvery longÕ drift (10-20 m) by a reduced 
electric field. 

\subsection{A low momenta charged particle test beam campaign}
The actual physics performance of new massive detectors
will directly depend their ability to reconstruct, measure and
identify low energy particles. In particular electron separation
from abundant $\pi^0$ background will be crucial.
It is mandatory to consider dedicated test beam campaigns, to test and optimize the readout
methods and to assess the calorimetric performance of liquid Argon detectors.
The proposed test beam will address the following points:
\begin{enumerate}

\item {\bf Electron, neutral pion, charged pion, muon reconstruction}: 
A crucial feature of the LAr TPC is a very fine sampling, 
which should deliver unmatched performances in particle identification and reconstruction. 
The reconstruction of electrons, neutral and charged pions and muons will be demonstrated. 
The obtained results will allow to further optimize the readout parameters for future large detectors.

\item {\bf Electron/$\pi^0$ separation}: A crucial feature of the LAr TPC is the possibility to precisely measure
and identify electrons from neutral pion backgrounds. 
The experimentally achievable separation will be demonstrated by inserting a hydrogenate target 
inside the detector in order to collect a significant sample of charge exchange events $\pi^- p\rightarrow n+\pi^0$.
The obtained results will allow to further optimize the readout parameters for future large detectors.

\item {\bf Calorimetry}: A specific feature of the LAr TPC is its 100\% homogeneity and full sampling capabilities. 
As an extension of the measurements performed in above, more refined measurements with low energy particles
 (0.5-5 GeV/c $e/\mu/\pi$) will yield actual calorimetric performance and determine the ability to reconstruct full neutrino events in 
 the GeV-range. These results will play an important role in future projects involving low energy neutrino beams or 
 sensitive searches for proton decay and complement direct measurements in a low energy neutrino beam.
 
\item {\bf Hadronic secondary interactions}: If sufficient statistics is collected, an exclusive final state study of 
pion secondary interactions will be attempted. Comparison of the data obtained with MC (e.g. GEANT4) will
allow to cross-check and eventually tune these models. These results are relevant for running experiments
(e.g. T2K).

\item {\bf Stopping muons and pions:} If sufficient low momenta are achieved, then a sample of charge-selected
stopping muons and pions will be studied in order to assess the imaging of the decay chains (for positive
particles) and of the capture (for negative particles).
 


\end{enumerate}

\subsection{A 1 kton full engineering demonstrator acting as near detector for neutrino fluxes and cross-section measurements}

Very few real neutrino events in liquid Argon have been fully reconstructed to allow a complete understanding
of the complex physics and detector effects at play. The only well-studied sample comes from 
a small 50~lt chamber developed by the ICARUS Collab.
and exposed to the CERN WANF high energy neutrino beam which collected
less than 100~quasi-elastic events~\cite{Arneodo:2006ug}.
We show below a list of effects that need to be studied~\cite{Badertscher:2008bp}:

\begin{itemize}
\item Neutrino interaction:
      \begin{itemize}
       \item Fermi motion and nuclear binding energy,
       \item Nuclear interactions of final state particles within the hit nucleus (FSI),
       \item Vertex nuclear remnant effects (e.g. nuclear break-up signal),
       \item Neutral Current (NC) $\pi^{0}$ event shape including vertex activity.           
      \end{itemize}
\item Detector medium:
      \begin{itemize}
       \item Ionization processes,
       \item Scintillation processes,
       \item Correlation of between amount of charge and light,
       \item Charge and light quenching,
       \item Hadron transport in Argon and secondary interactions,
       \item Charge diffusion and attenuation due to impurity attachment.
      \end{itemize}      
\item Readout system including electronics system:
      \begin{itemize}
       \item signal amplification or lack thereof,
       \item signal-to-noise ratio,
       \item signal shaping and feature extraction.
      \end{itemize}       
\item Reconstruction:
      \begin{itemize}
       \item Pattern recognition
       \item Background processes (NC $\pi^0$, $\nu_\mu$~CC, ...) and their event shape
       \item Particle identification efficiency and purity
      \end{itemize}       
\end{itemize}

The reconstruction of electromagnetic showers with  $\pi^0$  decay was performed on the ICARUS T300
data collected on surface~\cite{:2008sz} yielding 196 candidates with a mean energy of $\sim 700~$MeV.
During the CNGS runs the ICARUS T600 at LNGS will collect about 1300~$\numu$~CC (400~$\numu$~NC) per $4.5\times 10^{19}$~pots 
(=1 nominal year). However, only about 10\% of these events will be in the relevant kinematical region 
relevant to future neutrino oscillation experiment owing to the very high CNGS beam energy.

Almost all physics performance studies for future long-baseline experiment using liquid Argon TPC
rely on a single Monte-Carlo study of $\pi^0$ rejection~\cite{yuange}. The sensitivity to $\sin^22\theta_{13}$
will directly depend on the actual $\pi^0$ rejection performance achieved with real data.

It is clear that the energy resolution and physics performance will depend on several detector
parameters, including the readout pitch, the readout method chosen and
on the resulting signal-over-noise ratio ultimately affecting the reconstruction of the events.
We therefore stress that significantly
improved experimental studies with prototypes exposed to
neutrino beams of the relevant energies and sufficient statistics 
are mandatory to assess and understand these effects.
We mention three configurations that have been proposed in this
context:
\begin{enumerate}
\item a 100~ton LAr TPC was proposed for the T2K 2~km site~\cite{loi2km};
\item a 250~kg LAr TPC ArgoNEUT ~\cite{argoneut}
has been installed in front of the MINOS near detector and 
will be taking data soon;
\item the MicroBOONE experiment~\cite{microboone} aimed at testing
the MiniBoone anomaly, is in the design phase. 
\end{enumerate}
From these three efforts, the ArgoNEUT is the most advanced (although smallest)
and is expected to provide results within the end of 2009. MicroBOONE has received stage-1 approval
and is planned to take data in FY2011.

Beyond these efforts, we believe that a 1~kton scale device is the appropriate choice
for a full engineering prototype of a 100~kton detector. The chosen size for the prototype is the result of
two a priori contradictory constaints: (1) the largest possible detector as to minimize
the extrapolation to 100~kton (2) the smallest detector to minimize timescale
of realisation and costs. A 1~kton detector can be built assuming the GLACIER design
with a 12~m diameter and 10~m vertical drift. From the point of view of the drift path,
a mere factor 2 will be needed to extrapolate from the prototype to the 100~kton device.
Hence, the prototype will be the real demonstrator for the long drifts. At the same time,
the rest of the volume scaling from the 1~kon to the 100~kton achieved by increasing the diameter to about 70~m,
can be realized noting that (a) large LNG tanks with similar diameters
and aspect ratios already exist (b) the LAr LEM-TPC readout above the liquid will
be scaled from an area of 80~m$^2$ (1 kton) to 3800~m$^2$ (100 kton). This will not
require a fundamental extrapolation of the principle, but rather 
only pose technical challenges of production.

\section{Conclusions}
Two generations of large water Cherenkov detectors at Kamioka (Kamiokande and Super-Kamiokande) have been extremely successful. 
And there are good reasons to consider a third generation water Cherenkov detector with an order of magnitude larger mass than Super-Kamiokande 
for both non-accelerator (proton decay, supernovae, ...) and accelerator-based physics.
In parallel, the pioneering developments of the ICARUS liquid Argon TPC with immersed readout wires, already the fruit of several decades of 
R\&D, have not yet led to detectors competing with Super-Kamiokande, in spite of significant advances and better intrinsic 
physics performance. At this point in time the realization of the ultimate LAr TPC that can compete with the planned third 
generation water Cerenkov detectors, offers great promises but also many challenges. 
Only a very massive underground liquid Argon detector of about 100 kton could represent a credible alternative.

We have described the GLACIER concept:
its design, based on the LAr LEM-TPC, represents a scalable, 
cost-effective LAr detector up to possibly 100 kton.
We have already engaged on a graded R\&D strategy towards the GLACIER detector.
We plan to assess the physics performance of LAr detectors 
with test beam campaings.
The LAGUNA design study will define the optimal underground detector sites
for such future very large neutrino detectors. 
ArDM-1t is a real 1-ton scale prototype of the GLACIER concepts. ArgonTube will be a dedicated measurement of long drifts.
After a successful completion of these steps we want to proceed to a proposal for a 100 kton-scale underground detector, 
which would include the discussion of a 1~kton full engineering prototype.	

In all cases, we must wait for the results
from T2K and DOUBLE-CHOOZ to define the next major step. 
If a positive evidence
for $\theta_{13}$ is found, a tremendous momentum will be created towards
the discovery of CP-violation in the lepton sector.
In this context, the CNGS and JPARC conventional beams will be providing neutrinos in the coming years,
offering the opportunity to address the necessity and feasibility of upgrades. Most probably
CNGS will have to be substantially upgraded to remain competitive
with JPARC. If this is not practical, a new CERN superbeam, optimized for CP-violation
and directed towards one of the seven LAGUNA sites should be considered.
In absence of a positive result from T2K and DOUBLE-CHOOZ, large underground detectors 
could be operated as well in connection with other, more advanced neutrino beams 
like for instance beta-beams or neutrino factories.

\section*{Acknowledgments}
I would like to thank the organizers of DICRETEÕ08 for their kind invitation, 
warm hospitality, great atmosphere and excellent scientiÞc level of the meeting. 
I also acknowledge fruitful discussions with many colleagues, in
particular Takuya Hasegawa, 
Alberto Marchionni, Takasumi Maruyama, Anselmo Meregaglia,
Filippo Resnati, and Christos Touramanis, and also Dick Gurney
from Technodyne.


\end{document}